\documentclass[11pt,preprint]{aastex}

\newcommand{\fnu}{$f_{\nu}$}

\shorttitle{Radio Emission from Late-M and L Dwarfs}
\shortauthors{Burgasser \& Putman}

\begin{document}

\title{Quiescent Radio Emission from Southern Late-type M and L Dwarfs and a
Spectacular Radio Flare from the M8 Dwarf DENIS 1048$-$3956}

\author{Adam J.\ Burgasser\altaffilmark{1}}
\affil{Department of Astrophysics, American Museum of Natural History,
Central Park West at 79th Street, New York, NY 10024 \email{adam@amnh.org}}

\and

\author{Mary E.\ Putman}
\affil{Department of Astronomy,
University of Michigan,
Ann Arbor, MI 48109-1090
\email{mputman@umich.edu}}

\altaffiltext{1}{Spitzer Fellow}

\begin{abstract}
We report the results of a radio monitoring program conducted at
the Australia Telescope Compact Array to search for quiescent and
flaring emission from seven nearby Southern late-type M and L
dwarfs. Two late-type M dwarfs, the M7 V LHS 3003 and the M8 V DENIS 1048$-$3956,
were detected in quiescent emission at 4.80 GHz. The
observed emission is consistent with optically thin
gyrosynchrotron emission from mildly relativistic ($\sim$1--10
keV) electrons with source densities $n_e \lesssim 10^{9}$
cm$^{-3}$ in $B \gtrsim 10$ G magnetic fields.
DENIS 1048$-$3956 was also detected in two spectacular, short-lived
flares, one at 4.80 GHz (peak {\fnu} = 6.0$\pm$0.8 mJy) and one at
8.64 GHz (peak {\fnu} = 29.6$\pm$1.0 mJy) approximately 10 minutes
later.  The high brightness temperature (T$_B \gtrsim 10^{13}$ K),
short emission period ($\sim$4-5 minutes), high circular
polarization ($\sim$100\%), and apparently narrow spectral
bandwidth of these events imply a coherent emission process in a
region of high electron density ($n_e \sim 10^{11}$-$10^{12}$
cm$^{-3}$) and magnetic field strength ($B \sim 1$ kG). If the two
flare events are related, the apparent frequency drift in the emission suggests
that the emitting source either moved into regions of higher
electron or magnetic flux density; or was compressed, e.g., by
twisting field lines or gas motions.  This emission may be related to a recent
optical flare from this source that exhibited indications of
chromospheric mass motion. The quiescent fluxes
from the radio-emitting M dwarfs are too bright to support the G\"{u}del-Benz
empirical radio/X-ray relations, confirming a trend previously noted
by Berger et al.  The violation of these relations is
symptomatic of a divergence in magnetic emission trends at and beyond
spectral type M7/M8, where relative X-ray and H$\alpha$ emission
drops precipitously while relative radio emission appears to
remain constant or possibly increases.  With an apparent decline
in chromospheric/coronal heating, the origin of hot coronal
plasmas around ultracool dwarfs remains uncertain, although
external sources, such as accretion from a residual disk or
tidally distorted companions, remain possibilities worth
exploring.
\end{abstract}

\keywords{stars: activity --- stars: flare --- stars: low-mass, brown dwarfs --- stars: individual (DENIS J104814.7-395606, LHS 102B, LHS 3003) --- techniques: interferometric --- radio continuum: stars}

\section{Introduction}

Magnetic fields are fundamental to stars, playing an important
role in early accretion, angular momentum evolution, and a
number of interaction mechanisms.
The presence and strength of magnetic fields above the surface of a cool star,
when not directly measured from Zeeman line broadening (e.g., Saar \& Linsky 1985,
Johns-Krull \& Valenti 1996),
are generally inferred
from the presence of high temperature coronal and chromospheric emission from radio
to X-ray wavelengths.  This high-temperature emission, which may be characterized
as quiescent (stable or slowly varying) or flaring (rapid variation, high energy),
must arise from nonradiative heating processes, for which magnetic activity
is regarded as the most probable source. Magnetic activity is common amongst
M-type stars, with the frequency and strength of quiescent
H$\alpha$ emission, indicating the presence of a hot chromosphere,
peaking around spectral type M7/M8 \citep{giz00,wes04}.  For even
cooler stars and brown dwarfs, including ultracool late-type M,
L, and T dwarfs \citep{kir99,me02a},
H$\alpha$ emission declines rapidly,
both in strength and frequency, so that few
field objects later than type L5 exhibit any optical emission whatsoever \citep{giz00,me02b}.
Similar trends are also found in quiescent X-ray emission \citep{neu99,fle03}.

The reduction of chromospheric and coronal emission in ultracool
dwarfs is broadly consistent with theoretical expectations.  The
cool, dense atmospheres of these objects imply low ionization
fractions and thus high electric resistivities, so that magnetic
field lines are largely decoupled from the upper atmosphere
\citep{mey99,moh02,gel02}. As a result, the generation and
propagation of magnetic stresses, which can lead to magnetic
reconnection followed by electron/ion acceleration, is inhibited
at the stellar surface. The generation of the magnetic field
itself may also be impeded, as the fully convective interiors of
objects with masses M $\lesssim$ 0.3 M$_{\sun}$ \citep{cha97}
inhibits the standard $\alpha$$\Omega$ dynamo mechanism, although
other dynamo mechanisms --- turbulent \citep{dur93} or
${\alpha}^2$ dynamos \citep{rad90} --- may be sufficient.

The occurrence of flaring emission does not appear to drop off as rapidly as quiescent
emission, as objects as late as spectral type L5 \citep[T$_{eff} \approx 1700$ K]{hal02a,hal02b,giz02,lie03},
and possibly T6 \citep[T$_{eff} \approx 1000$ K]{me00b,me02b}, have been detected in outburst.
Optical flares, both line and continuum emission, typically last on order hours
and have duty cycles (percentage of time in strong emission)
of $\lesssim$1--7\% \citep{rei99,giz00,lie03}.  These flares
have been detected on objects
with no observable quiescent emission \citep{bas95,rei99}.
X-ray flares have been detected in field objects down to spectral types M9/M9.5 \citep{fle00,rut00,scm02,ham04}.
The presence of flaring, high-energy emission
from very late-type dwarfs is strong evidence that magnetic fields are
present.
However, based on the theoretical expectations outlined above, it still remains unclear
how magnetic flaring energy is transmitted through largely neutral atmospheres \citep{moh02}.

Given these activity trends near the M/L dwarf boundary, nonthermal magnetic radio
emission from cool dwarf stars and brown dwarfs has largely been assumed to be exceedingly weak.
This expectation is supported by the G\"{u}del-Benz
relations \citep[hereafter GB relations]{gud93,ben94}, an empirical correlation between
radio and X-ray luminosities that holds over several orders of magnitude over much
of the low-mass stellar main sequence and for interacting binary/active systems.
The GB relations predict radio fluxes
$f_{\nu} \lesssim$ 1 $\mu$Jy for nearby ultracool M dwarfs \citep[hereafter B02]{brg02},
undetectable with current instrumentation.
Early radio studies of the latest-type dwarfs appeared to confirm
this prediction (e.g., Krishnamurthi, Leto \& Linsky 1999).
However, \citet[hereafter B01]{brg01} detected
both quiescent and flaring radio emission from the 500 Myr M9 brown dwarf LP 944-20,
an object that is both very cool
(T$_{eff}$ $\approx$ 2150 K; Dahn et al.\ 2002) and undetected in quiescent X-rays \citep{rut00}.
This radio emission violates the GB relations by roughly four orders of magnitude.
Three other late-type dwarfs spanning M8.5 to L3.5 were subsequently
detected in quiescent and flaring radio emission (B02),
at levels much higher than expected from the GB relations.
These observations indicate that both magnetic fields and sustained coronal
plasmas are present
above the photospheres of ultracool dwarfs at and below the Hydrogen burning limit,
and point to a fundamental change in the magnetic emission process at
low T$_{eff}$.

To further explore nonthermal emission from ultracool dwarfs, we have used the
Australia Telescope Compact Array (ATCA) to search for radio emission from
seven nearby, late-type M and L dwarfs in the Southern hemisphere.
In $\S$ 2 we describe the sample, observations, and data reduction procedures.  Detection
of quiescent emission from two late-type M dwarfs are discussed and analyzed in $\S$ 3.  One of
the quiescent sources, the M8 dwarf DENIS 1048$-$3956, was also detected
in two short, powerful flares in
both of the frequency bands observed; this emission is discussed in $\S$ 4.  In $\S$ 5
we examine possible trends of radio
emission with spectral type, rotation, and the existence of optical quiescent and/or flaring emission;
and speculate on the origin of coronal plasma above cool photospheres.
Results are summarized in $\S$ 6.

\section{Observations}

\subsection{The Sample}

Seven nearby ultracool dwarfs in the Southern hemisphere were selected
for observation; their properties
are summarized in Table 1.  The primary selection criteria
were (1) spectral type M7 or later, where the GB relations appear
to break down (B02); and (2) proximity to the Sun.  As such, our sample
spans the spectral type range M7 to L5 and
have distance measurements or spectrophotometric estimates
of 11 pc or closer (with the exception of 2MASS 1139$-$3159, see below).
Detailed descriptions of the targets are as follows:

\noindent{\it LHS 102B:} Identified by \citet{gld99}, this object is a common proper motion companion
to the M3.5 V high-proper motion star
LHS 102 (a.k.a.\ GJ 1001), which has a parallactic distance measurement
of 9.55$\pm$0.10 pc \citep{van95}. Its L5 spectral type
suggests that it is cool enough
to be substellar (T$_{eff} \approx 1800$ K;
Leggett et al.\ 2002), although the absence of Li I absorption at 6708 {\AA}
implies M $>$ 0.06 M$_{\sun}$ \citep{reb92}.
\citet{gld99} estimate M = 0.072 M$_{\sun}$ for an age of 5 Gyr; i.e., at the Hydrogen burning limit.
High-resolution spectroscopy by \citet{bas00} shows that LHS 102B is a rapid
rotator, with $v\sin{i} = 32.5{\pm}2.5$ km s$^{-1}$.  Weak H$\alpha$ emission is
also seen in its optical spectrum.  \citet{gol04} have recently resolved this source as a
0$\farcs$086 (0.8 AU), equal-mass binary.

\noindent{\it SSSPM 0109$-$5100:} Identified by \citet{lod02} in the SuperCOSMOS Sky Survey
\citep{ham01}, this object has a near-infrared spectrum consistent with an L2 dwarf.
\citet{sch02} estimate a distance of $\sim$13 pc
based on its photographic $R$ and $I$ magnitudes and spectral type.  Using the $M_J$/spectral type
relation of \citet{cru03} and 2MASS photometry \citep{cut03},
we estimate a distance of $\sim$10 pc.  With no published
optical spectrum available, it is unknown as to whether this source
has quiescent H$\alpha$ emission or Li I absorption.

\noindent{\it 2MASS 0835$-$0819:} Identified and optically classified by \citet{cru03}, this
L5 dwarf has an estimated T$_{eff} \sim 1700$ K, based on the temperature/spectral type
relation of \citet{gol04}.  Like LHS 102B, 2MASS 0835$-$0819 is likely at or below the substellar limit.
Li I absorption
is not seen in its low-resolution optical spectrum, however, nor is
quiescent H$\alpha$ emission.  \citet{cru03} estimate the distance of
2MASS 0835$-$0819 at $\sim$8 pc.

\noindent{\it DENIS 1048$-$3956:} Identified in the DENIS survey \citep{epc97} by
\citet{del01}, this bright source ($J = 9.54{\pm}0.04$) has a high proper motion,
$\mu = 1{\farcs}529{\pm}0{\farcs}017$ yr$^{-1}$.
\citet{neu02} measure a parallactic distance of $4.6{\pm}0.3$ pc (see
also Deacon \& Hambly 2001), making this the closest star in our sample.
Originally classified M9 by \citet{del01}, we adopt the revised classification
of M8 from \citet{giz02}.
High resolution optical spectroscopy by \citet{fuh04} indicate $v\sin{i} = 25{\pm}2$ km s$^{-1}$,
consistent with measurements by \citet{del01}, making DENIS 1048$-$3956 another rapid rotator.
Quiescent and variable H$\alpha$ emission has been detected from this object \citep{del01,neu02,giz02},
while \citet{fuh04} have detected a massive optical flare, including blueshifted
components indicative of mass motion.  \citet{scm04} report an X-ray luminosity upper
limit of $L_X < 2{\times}10^{26}$ erg s$^{-1}$ based on the absence of this source
in the R\"{o}ntgen Satellite (ROSAT) all-sky survey catalog.

\noindent{\it 2MASS 1139$-$3159:} Identified by \citet{giz02}, this M8 dwarf
is the only object in our sample with a spectrophotometric distance beyond 11 pc.
2MASS 1139$-$3159 was chosen for its possible
membership in the $\sim$ 10 Myr TW Hydra Association
\citep{del89,kas97}, which would make it a young, very low-mass (M $\sim$ 0.025 M$_{\sun}$)
brown dwarf.  Optical spectroscopy from \citet{giz02} shows both H$\alpha$ and He I (6679 {\AA})
emission, along with low surface gravity features
indicative of a young, low-mass brown dwarf.
Li I absorption has not been reported, however.

\noindent{\it LHS 3003:} With a parallactic distance of 6.56$\pm$0.15 pc \citep{ian95},
this M7 dwarf is a nearby and well-studied system.
LHS 3003 was originally identified as a cool star
by \citet{rui90}, who observed a full sequence of Balmer H I emission while this
object was in a flare state.  Quiescent H$\alpha$ emission has also been observed
at the level $\log{({\rm L}_{H{\alpha}}/{\rm L}_{bol})} \approx -4.3$
\citep{tin98,moh03}.  In addition, ROSAT observations by
\citet{sch95} detected this object in soft X-rays (0.1--2.4 keV)
at the level of L$_X \approx 2{\times}10^{26}$ erg s$^{-1}$, or
$\log{({\rm L}_X/{\rm L}_{bol})} \approx -4.0$.
High resolution optical spectroscopy by \citet{moh03} indicate that this source
is a slow rotator, with $v\sin{i} = 8.0{\pm}2.5$ km s$^{-1}$.

\noindent{\it 2MASS 1534$-$1418:} Identified by \citet{giz02}, this M8 dwarf has
a spectrophotometric distance of $\sim$11 pc.  Quiescent H$\alpha$ emission is
seen in its low-resolution optical spectrum, but there has been
no additional follow-up of this source published in the literature.

\subsection{Data Acquisition and Reduction}

All observations were conducted with ATCA in its fully extended 6A configuration
(baselines of 0.63--5.94 km) during two runs on 2002
May 16-17 and 2002 Nov 29-Dec 2 (UT).   A log of observations is given in Table 2.
Sources were tracked in continuum mode simultaneously at 4.80 and 8.64 GHz
(6 and 3 cm) using
the broadest bandwidth available (128 MHz over 32 channels, binned to 13 independent channels) and
sampling every 10 s.
Nearby secondary calibrators selected from the ATCA Calibrator Catalog\footnote{See
\url{http://www.narrabri.atnf.csiro.au/calibrators/c007/atcat.html}.} were interspersed every 30-45 minutes
for relative flux and phase correction, and the primary calibrators PKS B0823-500 and PKS B1934-638 were
observed for absolute calibration at the beginning and/or end of each target cycle.
Sources were tracked for 10-12 hr depending on the declination, with on-source times of
roughly 8-10 hr each.

Visibility data were reduced in the MIRIAD
environment\footnote{See \url{http://www.atnf.csiro.au/computing/software/miriad/index.html}.}
using standard routines.  First, poor baselines in the target and calibrator
sources were flagged by visual inspection, both before and after phase and flux
calibration, by checking antenna leakage ($\lesssim$ 1\%), phase and flux stability of
primary and secondary calibrators,
and secondary calibrator polarization ($\lesssim$ 3\%).  Phase and flux calibration
of the target observations were tied to the secondary calibrators, which were in turn
tied to the primary calibrators.
The fully calibrated visibility datasets were then inverted and cleaned using the MIRIAD
routines {\it INVERT}, {\it CLEAN}, and {\it RESTORE} to produce
imaging data for source verification and measurement.
Radio fluxes were measured using the
{\it IMFIT} routine, while uncertainties were estimated
from the standard deviation of the imaging
data over a ${\sim}2{\arcmin}{\times}2{\arcmin}$ area without sources near the target position.
For DENIS 1048$-$3956, these uncertainties are slightly higher than expected
due to sidelobes from the bright radio source NVSS 104748$-$395053
(Condon et al. 1998; $f_{1.4GHz} = 120{\pm}4$ mJy),
7$\farcm$2 northwest of the target.
Because of its complex double-lobed morphology,
we were unable to model and subtract this background source from
the visibility data.  However, its influence in
the region of DENIS 1048$-$3956 is minimal (Figure 2), and the source
was not present in the 8.64 GHz band nor in the Stokes Q, U, or V
polarization images.

For time series data ($\S$ 4), visibilities
for each polarization (Stokes I, Q, U, and V) were independently averaged
across all baselines to measure the total radio flux, and monochromatic
flux densities were computed by averaging the central nine channels (${\Delta}{\nu} = 72$ MHz)
in the frequency domain.  Uncertainties in the time series data
were estimated from the standard
deviation of the averaged visibilities over 30 min intervals (in the absence of
flaring emission); i.e., assuming slow variation in the total source and background
radio emission in each field.  These uncertainties were typically of order 1 mJy per 10 s
time bin.

\section{Quiescent Emission}

\subsection{Detections}

For our targeted observations,
we adoped a somewhat less stringent 3$\sigma$ limit (0.10--0.12 mJy) for source detection
than the typical 4-5$\sigma$ limits used for survey work (e.g., Richards et al.\ 1998).
Only two of our targets had spatially coincident
quiescent radio sources
above this threshold in the 4.80 GHz band,
LHS 3003 and DENIS 1048$-$3956.
Imaging data for these two sources in
the Stokes I polarization at both frequencies are shown in Figures 1 and 2.
Note that the images for DENIS 1048$-$3956 do not include visibility data during the
periods of flaring observed from this object ($\S$ 4).
The radio flux peaks detected near these sources are found to be
within the mean beam size of the
predicted positions of the targets as determined from 2MASS astrometry
(accurate to within 0$\farcs$3; Cutri et al.\ 2003)
and proper motion measurements from the literature \citep{tin96,neu02}.
The relatively bright source ($f_{\nu} = 0.27{\pm}0.04$ mJy)
coincident with LHS 3003 has a faint $\sim$5$\arcsec$ extension
toward the northeast which also appears in the 8.64 GHz image.
We cannot rule out noise as the origin of this extended emission.
The fainter source ($f_{\nu} = 0.14{\pm}0.04$ mJy) coincident with
DENIS 1048$-$3956 has a shape consistent with the beam profile.
No significant, spatially coincident radio sources were found at
8.64 GHz for any of the targets.  All measurements are given in Table 3.

At the faint flux levels probed by our observations, background confusion is an
important consideration.  We therefore estimated the probability that the
detected 4.80 GHz sources are associated with LHS 3003 and DENIS 1048$-$3956
by computing the expected number of background sources ($N$) with similar
brightnesses present within the ATCA beam.  A 6 cm Very Large Array (VLA) survey of the Lockman Hole by
\citet{cil03} identified 28 sources (corrected for completeness to 28.6 sources)
with $f_{\nu} >$ 0.113 mJy in a 0.087 deg$^2$ area, implying an integrated source
density $N \approx 2.5{\times}10^{-5}$ arcsec$^{-2}$ for $0.1 \lesssim f_{\nu} \lesssim$ 20 mJy.
This is consistent with results from other deep 6 cm surveys \citep{alt86,don87,fom91}.
Based on the beam sizes listed in Table 2, this background density implies
a confusion probability $1-e^{-N} \lesssim$ 0.3\% and 0.2\% for LHS 3003 and DENIS 1048$-$3956,
respectively, ruling out confusion with high confidence.
We therefore conclude that quiescent emission from LHS 3003 and DENIS 1048$-$3956
at 4.80 GHz was detected.

Examination of the Stokes Q, U and V 4.80 GHz images for LHS 3003 and DENIS 1048$-$3956
show no significant sources.  However, these non-detections give only
weak constraints on the polarization of the quiescent emission.
Circular polarization upper limits (3$\sigma$) at 4.80 GHz are
$\Pi_V \equiv$ V/I $< 44$\% and $<$ 86\% for LHS 3003 and DENIS 1048$-$3956, respectively.

\subsection{Characterizing the Quiescent Emission}

The flux densities of the two detected M dwarfs imply frequency-dependent
radio luminosities
$L_{\nu,q} \equiv 4{\pi}f_{\nu}d^2 \approx (4-13){\times}10^{12}$
erg s$^{-1}$ Hz$^{-1}$
(Table 3), where $d$ is the distance to the source.
These values are similar to measurements for hotter M stars as well as many of the late-type
dwarfs detected by B02, although most of those detections were made at 8.46 GHz.
The non-detections in our sample generally have
luminosity upper limits brighter than the detections.

The brightness temperature of the radio emission at frequency $\nu$,
\begin{equation}
{\rm T}_B = 2{\times}10^9 (f_{\nu}/{\rm mJy}) ({\nu}/{\rm GHz})^{-2} (d/{\rm pc})^2 ({\cal L}/{\rm R}_{Jup})^{-2}~~{\rm K},
\end{equation}
\citep{dul85} provides a measure of the energetics of the
emitting electron population.  ${\cal L}$ is the length scale of the emitting region,
normalized here to the typical radii of very low mass
stars and brown dwarfs, R$_*$ $\sim$ 0.1 R$_{\sun}$ $\sim$ 1 R$_{Jup}$ $\approx 7{\times}10^9$ cm
\citep{bur01}. Assuming M-type stellar coronal dimensions,
${\cal L} \sim (2-3)\times$R$_*$ \citep{ben95}, the detected radio emissions
imply T$_B \approx (3-30){\times}10^7$ K (Table 3).  The temperature
of the emitting electrons, T$_e$, is related to the brightness
temperature by T$_e$ = T$_B$ for optically thick emission, and
T$_e$ = $\tau_{\nu}$T$_B$ for optically thin emission, where
$\tau_{\nu}$ is the frequency-dependent optical depth of emission.  The absence
of emission at 8.64 GHz for any of these sources implies
that the quiescent flux peaks near or below 4.80 GHz, so that
T$_e \lesssim$ T$_B$ $\sim$ 10$^7$-10$^8$ K.
These values are similar to coronal (ion) plasma temperatures of other late-type M dwarfs
derived from X-ray measurements \citep{gia96,rut00,fei02,fle03}.
Note that a more extended corona, such as that proposed by \citet{fle03} for the M8 dwarf VB 10
(${\cal L} \lesssim 20 {\rm R}_*$) would imply brightness temperatures that are significantly
lower.  On the other hand, VLBI measurements of the M-type
flare stars EQ Peg B and AD Leo find
${\cal L} \lesssim 2 {\rm R}_*$ \citep{ben95,let00}.
For lack of further observational constraints, we assume the source scale used above.

The inferred brightness temperatures imply a population of
mildly relativistic (1-10 keV) electrons in the emitting region.
Hence, gyrosynchrotron
emission is likely the source of the observed quiescent flux, a mechanism commonly
prescribed for persistent emission from late-type stars \citep{gud02}.
We can estimate the total radio luminosity of each source by
assuming emission below a peak frequency, $\nu_{pk}$, scales as ${\nu}^{2.5}$,
and emission above ${\nu}_{pk}$ scales as ${\nu}^{\alpha}$, where
$\alpha = 1.22-0.9\delta$ for a power-law electron distribution
$n(E) \propto E^{-\delta}$ \citep{dul82,dul85}.  Typical
coronal values of $\delta \approx 2-4$ \citep{gud02} imply $\alpha \approx -1.5$, consistent
with our 8.64 GHz upper limits ($\alpha < -0.4$ and $-1.2$).
Assuming ${\nu}_{pk} \approx 5$ GHz and emission over a harmonic range of 100
(${\nu}_{pk}/10 < \nu < 10{\nu}_{pk}$),
we estimate $L_R = \int{L_{\nu,q}d\nu} \approx (3-10){\times}10^{22}$ ergs s$^{-1}$,
or $\log{L_R/L_{bol}} \approx -7.3$ and $-7.7$ for LHS 3003 and DENIS 1048$-$3956, respectively.
These values are similar to those
obtained by B02 for their late-type M dwarf quiescent detections.

For gyrosynchrotron emission, the peak frequency of the radio flux for a power-law
electron distribution is related
to the electron density ($n_e$), length scale and magnetic field strength ($B$) of the emitting region
as\footnote{Note that the relations for gyrosynchrotron emission given in \citet{dul85} and \citet{gud02}
assume a lower electron energy cutoff of 10 keV, whereas we estimate energies down
to 1 keV for our sources.  Derivation of accurate expressions in the lower energy regime
are beyond the scope of this paper, and we assume our estimates to be accurate at the
order-of-magnitude level.}
\begin{equation}
\nu_{pk} \approx 16.6 n_e^{0.23}{\cal L}^{0.23}B^{0.77}~~{\rm kHz}
\end{equation}
\citep{dul85}, where we have assumed $\delta \sim 3$ and an average pitch angle
$\theta = {\pi}/3$ \citep{gud02}.  Using the length scale above and again assuming ${\nu}_{pk} \approx$ 5 GHz,
Eqn.\ 2 reduces to $B \approx 11n_9^{-0.3}$, where $n_9 = n_e/(10^9$ cm$^{-3}$).
We can further use the requirement that Razin-Tsytovich suppression
\citep{tsy51,raz60}
is minimal at the frequencies observed, implying that emission occurs above a minimum
frequency $\nu_{min}$, and hence
\begin{equation}
\nu_{pk} \gtrsim \nu_{min} \gtrsim \frac{\nu_p^2}{\nu_c} \approx 29\frac{n_9}{B}~~{\rm GHz}
\end{equation}
for mildly relativistic electrons
\citep{dul85}. Here, $\nu_p \equiv ({n_ee^2}/{\pi}m_e)^{1/2} \approx 0.28\sqrt{n_9}$ GHz
is the fundamental plasma frequency and ${\nu}_c = eB/2{\pi}m_ec \approx 2.8B$ MHz
is the cyclotron frequency.
Combining Eqns.\ 2 and 3 yields $n_e \lesssim 2{\times}10^9$ cm$^{-3}$ and $B \gtrsim$ 10 G
for both LHS 3003 and DENIS 1048$-$3956.

The quiescent magnetic field estimates, likely
accurate only to within an order of magnitude, are
roughly in agreement with those of B02 for their ultracool
dwarf detections.
Our electron density estimates, on the other hand, are $\sim$10$^3$ times smaller.
{\em Chandra} and XMM-Newton grating observations of the M3.5 V AD Leo yield
coronal electron density upper limits more consistent with our estimates,
$n_e \lesssim 10^{10}-10^{11}$ cm$^{-3}$ \citep{van03},
although the structure of the coronal region
of this star may be quite different than that of our cooler sources.

\section{Flaring Emission from DENIS 1048$-$3956}

\subsection{Detection and Characterization of the Flares}

Time series analysis of all targets was performed to search for variability
and flare events.  Only one source was
detected above our $\sim$3 mJy sensitivity threshold (3$\sigma$ standard deviation in 30 s binned
visibilities),
the quiescent emitter DENIS 1048$-$3956.  As shown in Figure 3, this
object underwent two strong and rapid flares, one each at 4.80 and 8.64 GHz.
Flaring emission is particularly strong
at 8.64 GHz, and can be seen in the visibility data.
These flares were very short-lived, with durations of $\tau \sim 4-5$ minutes
in each band; and occurred roughly 10 minutes apart,
peaking at 14:15:15 and 14:25:16
(UT) for the 4.80 and 8.64 GHz emission, respectively.
There was no significant emission in the opposite frequency for
either of the flares, indicating narrow-band emission.
Peak flux densities, $f_{\nu,f}$, estimated from
Gaussian fits to the unbinned time series data, were
6.0$\pm$0.8 mJy and 29.6$\pm$1.0 mJy at 4.80 and 8.64 GHz, respectively.
These fluxes are over ten times brighter than the radio flares observed on
LP 944-20 by B01, which lies at a similar distance as DENIS 1048$-$3956.
Finally, both flares were highly circularly polarized.  Figure 4 compares the Stokes Q, U, and V
polarizations for both flares.
Neither were detected in the Stokes Q and U polarizations.
Peak Stokes Q and U fluxes at 4.80 GHz have 1$\sigma$ upper limits of
1.1 and 1.2 mJy, respectively, implying peak polarizations
$\Pi_Q \equiv$ Q/I $<$ 18\% and $\Pi_U \equiv$ U/I $<$ 20\%.
At 8.64 GHz, $\Pi_Q$ and $\Pi_U$ are both $<$ 3\% (1$\sigma$).
On the other hand, Stokes V polarizations in both bands were essentially
unity at the peak of the flares (c.f., Figures 3 and 4).
Hence, essentially all of the emission is concentrated in circular
polarization.

Given the unique nature of these flares,
we carefully checked that emission originated from
the source by examining the visibility images over the entire observation period.  Distinct fringes,
expected from an unresolved point source observed with parallel baselines,
are seen with maximum intensity at the position of DENIS 1048$-$3956 in imaging data
that includes the flare period, and the fringes are spaced and oriented
in accordance with the alignment of the ATCA at the time of the flare emission.  The flares do
not arise from the bright radio source NVSS 104748$-$395053, as no flare emission is
found at the position of this source in the imaging data.   Indeed, the fringes
are most obvious in the Stokes V images in which there is no interfering
emission from NVSS 104748$-$395053.
We therefore conclude that the flaring radio
emission originates from DENIS 1048$-$3956 itself.

\subsection{Coherent Emission}

The strong, rapid, narrow-band, and highly polarized flares
detected from DENIS 1048$-$3956 are
quite different than those observed from late-type M and L dwarfs
by B02, which had longer flare periods ($\sim$6-25 min)
and somewhat less polarized ($\sim$30-66\%) emission.  Flares
detected on LP 944-20 by B01 exhibited coincident peaks at 4.86 and 8.46 GHz, as opposed
to the temporally offset flares seen here.
Furthermore, the brightness temperatures of the DENIS 1048$-$3956 flares are very high,
T$_B$ = $(1.1{\pm}0.2){\times}10^{10} ({\cal L}/{\rm R}_*)^{-2}$ and
$(1.7{\pm}0.2){\times}10^{10} ({\cal L}/{\rm R}_*)^{-2}$ K for the 4.80 and 8.64 GHz
flares, respectively.  Peak brightness temperatures for other radio flaring
late-type M and L dwarfs are a factor of 10 or more less.  On the other hand,
the DENIS 1048$-$3956 flaring emission is quite
similar to rapid ($\lesssim$10 min), highly polarized ($\gtrsim$60\%)
flares seen on earlier-type active M stars, including
the M5.5 V UV Cet A \citep{ben98,bin01}, the M4 V DO Cep \citep{whi89},
and AD Leo \citep{ste01}.  The last source
exhibited a rapid ($\sim$1 min) burst at 4.85 GHz with a
peak flux $f_{\nu,f} \approx$ 300 mJy,
T$_B \sim 5{\times}10^{10} ({\cal L}/{\rm R}_*)^{-2}$ K,
and nearly 100\% circular polarization, similar in scale and energetics to the emission seen
on DENIS 1048$-$3956.  \citet{ste01} argue that
the high temperature and polarization of this flare
is the result of a coherent emission process, as has been argued for other
rapid, high energy and high polarization flares \citep{bin01}.
The properties of the DENIS 1048$-$3956 flares indicate coherent emission
as well.

Two mechanisms are generally considered for coherent processes in cool stellar coronae: electron
cyclotron maser (ECM) and plasma emission.  Both produce
narrow bandwidth, highly polarized, and highly energized radio bursts.
The propagation of this emission is problematic at the frequencies observed
here, however, as free-free and gyroresonance absorption from ambient (thermal) elections will
suppress emergent radiation \citep{dul85}.  Indeed, coherent emission above 5 GHz is exceedingly
rare \citep{gud02}.  However, radiation can escape from regions with
sufficiently high density gradients.  The optical depth for free-free absorption is
\begin{equation}
{\tau}_{ff} \approx 15{\rm T}^{-3/2}(\nu/{\rm GHz})^2{\cal L}_n
\end{equation}
\citep{dul85}, where ${\cal L}_n \equiv n_e/{\nabla}n_e$ is the density gradient scale in cm.
For DENIS 1048$-$3956,
assuming that the ambient electron temperature is that derived from the quiescent emission,
$\tau_{ff}$ is less than unity for ${\cal L}_n \lesssim 3{\times}10^8$ cm $\sim 0.04$R$_*$,
implying T$_B \gtrsim 10^{13}$ K for ${\cal L} \sim {\cal L}_n$.
These values are similar to those derived for
the AD Leo flare, and again suggests similar emission mechanisms.

\citet{ste01} argue that plasma emission is more likely in the case of AD Leo
given the higher degree of gyroresonance absorption occurring
for ECM emission.  Plasma emission is peaked at the plasma frequency, implying
$n_e \approx 3{\times}10^{11}$ and $9{\times}10^{11}$ cm$^{-3}$
for the 4.80 and 8.64 GHz flares of DENIS 1048$-$3956, respectively.
Razin-Tsytovich suppression implies $B \lesssim 2-3$ kG.  On the other hand,
if ECM is responsible (as is argued for bursts on UV Cet; Bingham, Cairns, \& Kellett 2001),
emission at $\nu_c$ implies $B = 1.7-3.1$ kG and
$n_e \lesssim 10^{11}$ cm$^{-3}$ (constraining $\nu_p/\nu_c \lesssim 0.5$; Dulk 1985).
Hence, we estimate $B \sim 1$ kG and $n_e \sim 10^{11}$-$10^{12}$ cm$^{-3}$ in the
flaring region, similar
to values derived for the AD Leo flare \citep{ste01} but much
larger than estimates from the quiescent emission of this source.
The amplification of the magnetic field in the flaring region
could arise from twisted loops associated with magnetic reconnection sites.
The higher electron densities
are consistent with Solar flaring loop plasmas, which are 10--100 times
more dense than quiescent regions \citep{can90}.

\subsection{A Single Flare Event with Frequency Drift?}

The unique nature, temporal proximity, and non-simultaneous emission
during the DENIS 1048$-$3956 flares strongly suggests that
they are related.  We propose that the two flares are in
fact snapshots of a single narrow bandwidth flaring event whose source region evolved
in such a way as to cause a frequency drift in the emission.
Such narrow-band frequency drift has been previously observed in radio dynamic
spectra of active stars \citep{jac87,bas90}.  Assuming simplistically that
the propagation is linear in time, the temporal spacing of the emission
implies a drift rate $\dot{\nu} \approx +6.4$ MHz s$^{-1}$, of the same order as drift
rates observed from a 10 minute radio flare observed from the active M6 star
UV Cet \citep{jac87}. The similar (possible) drift rates and timescales suggest that
the emission from both stars could arise
from analogous processes.

Assuming for the purposes of discussion that a frequency drift is present, the shift toward
higher emission frequencies indicates that the source region evolved
toward conditions of higher magnetic field strength and/or electron density.
This transition could arise from physical movement of the source --- e.g.,
infalling into regions of higher density and/or field strength --- or
a modification of the source environment --- e.g., compaction of the emitting
region or a compression of field lines.
In either case, a frequency drift implies
$\dot{n_e} \approx +10^9$ cm$^{-3}$ s$^{-1}$
for plasma emission.  We can assign a drift or compaction velocity ($v_s$)
to the emitting region by assuming
${\cal L}_n \approx n_ev_s/\dot{n_e} \lesssim$ 0.04R$_*$,
so that $v_s \lesssim 5$ km s$^{-1}$. This value is similar to the velocity
of redshifted components seen in line emission from a massive optical flare
from DENIS 1048$-$3956 \citep{fuh04}.
While the long chain of assumptions used here cannot prove a connection
between the optical and radio flaring, the suggested agreement in the kinematics is intriguing.
It is also possible that the
radio flux is emerging from optically thick emission to optically thin
emission, consistent with the $\nu^{2.7}$ dependence between
the 4.80 and 8.64 GHz peak fluxes.  This emergence of the source region may
be the result of a clearing away of overlying absorbing plasma, possibly related to
the highly blueshifted ($v \sim 100$ km s$^{-1}$) components of the optical flare
detected by \citet{fuh04}.

The peak luminosities, $L_{\nu,f}$ from the flaring emission
are $1.5{\times}10^{14}$ and $7.5{\times}10^{14}$ erg s$^{-1}$ Hz$^{-1}$
for the 4.80 and 8.64 GHz spikes, respectively.  The total radio luminosity
depends on the frequency response of the emission.  At one extreme, if we assume that
the two flare events are
independent and confined to the observed frequency bands
($\Delta{\nu}$ = 72 MHz), then
$L_R \approx L_{\nu,f}{\Delta}{\nu} = 6{\times}10^{22}$ erg s$^{-1}$,
roughly equivalent to the persistent component.
On the other hand, if the flare emission is the result of a drifting source,
the emission band could be as broad as
${\Delta}{\nu} \approx \dot{\nu}{\tau} \approx 2$ GHz.
Assuming a Gaussian frequency distribution with a full width at half maximum
of 2 GHz, the equivalent radio luminosity is ten times greater, approaching 10$^{-6}L_{bol}$.
Similarly, the total energy released in
the flaring emission may range from 10$^{24}$ erg (observed emission) to
$>10^{26}$ erg for a drifting source.

\section{Discussion}

\subsection{Radio Emission Trends}

The detection of quiescent emission from a handful of
ultracool M and L dwarfs is
surprising in itself, but perhaps more interesting is that this emission may
in fact be common.
B02 found that the relative quiescent radio luminosity of their detected late-type sources,
$L_R/L_{bol}$, was constant or increasing with spectral type out to type L3.5.
This trend is contrary to the observed H$\alpha$
emission, which weakens rapidly beyond spectral type M7/M8 \citep{giz00,wes04};
and quiescent X-ray emission,
which appears to turn over around the same spectral types \citep{fle03}.
Figure 5 compares the ratios $L_{\nu,q}/L_{bol}$ and $L_{H\alpha}/L_{bol}$ versus spectral type
for field stars with spectral types
M2 to L6.  Values for $L_{\nu,q}$ at 3 or 6 cm were obtained from the
literature \citep[B01; B02]{lin83,whi89,gud93,kri99,let00}
and our own observations.  Bolometric luminosities
as a function of spectral type
were derived from a seventh order polynomial fit to empirical values
for M and L dwarfs in the 8 pc sample
\citep{rei00a} and from \citet{gol04}.  Values
for $L_{H\alpha}/L_{bol}$ are from \citet{haw96,giz00}, and \citet{me02b}.  The trend of
increasing relative radio luminosity is clearly apparent in this data,
extending well beyond the drop in H$\alpha$ emission.  A linear fit to radio detections
with spectral types M3 to M9 yields
\begin{equation}
\log{L_{\nu,q}/L_{bol}} = -18.1 + 0.11{\times}{\rm SpT},
\end{equation}
where SpT(M3) = 3, SpT(M9) = 9, etc.  This is similar to the relation
diagrammed in Figure 6b of B02.  Furthermore, the single L dwarf radio detection
(the L3.5 dwarf 2MASS 0036+1821) is consistent with an extrapolation of this trend.
One caveat, however, is that
many of the radio quiescent detections are close to the sensitivity limits of the observations.
Hence, non-detection upper limits are not strong constraints
for lower levels of emission (or non-emission), which may be orders of magnitude below this
line.  Nonetheless, with fourteen radio-emitting late-type M and L dwarfs within
12 pc of the Sun having detections or upper limits within 0.5 dex of this line,
there are strong indications of a general trend.

The few sources that have radio emission upper limits
below this line were closely examined by B02, who found that a dominant
fraction were slowly rotating ($v\sin{i} < 10$ km s$^{-1}$), late-type
M dwarfs.  This, argued B02, suggests a correlation between rotation and
radio emission analogous to the well-known activity-rotation relation for F-K
main sequence stars \citep{pal81,noy84}.
Again, our observations lend some support to this conclusion,
as DENIS 1048$-$3956 is clearly a
rapid rotator with $v\sin{i}$ = 25$\pm$2 km s$^{-1}$.
However, the brighter radio detection in our sample, LHS 3003,
has $v\sin{i} = 8.0{\pm}2.5$ km s$^{-1}$
\citep{moh03}, equivalent to the similarly-typed M7 dwarf VB 8 which has an upper
limit on its radio flux well below Eqn.\ 5 ($\log{L_{\nu,q}/L_{bol}} < -18.4$;
Krishnamurthi, Leto, \& Linsky 1999).  Of course, as $v\sin{i}$ provides
only a lower limit on the actual rotation velocity, it is possible
that LHS 3003 is a rapid rotator viewed close to
pole-on. However, the L5 2MASS 1507$-$1627,
which was undetected by B02 to a 3 cm limit of
$\log{L_{\nu,q}/L_{bol}} < -16.8$ (compared to $-16.5$ from Eqn.\ 6), is a
rapid rotator, with $v\sin{i} = 27{\pm}6$ km s$^{-1}$ \citep{bai04}.
Hence, rotation may not be the only factor driving radio emission.

Turning to the presence of H$\alpha$ emission, it is interesting to note that
both of our detected sources are quiescent H$\alpha$ emitters and therefore
have appreciable chromospheres.  On the other
hand, both 2MASS 1139$-$3159 and 2MASS 1534$-$1418, which
were not detected in the radio, also exhibit quiescent H$\alpha$ emission; while
two of the four sources detected by B02 (BRI 0021$-$0214 and 2MASS 0036+1821)
have little or no quiescent H$\alpha$ flux.  There is therefore no clear correlation of radio
coronal emission with optical chromospheric emission.  On the other hand, two of the four
late-type M and L dwarfs detected by B02 and both sources detected in our study
have been observed in strong H$\alpha$
or X-ray flaring emission.  This includes the rapidly-rotating
M9.5 dwarf BRI 0021$-$0214 which exhibits
no quiescent H$\alpha$ emission \citep{bas95,rei99}.
Since flaring emission is fairly
rare, it is possible that the other radio
detections are flare stars that have not yet been observed in optical or X-ray emission.
It is again important to consider contrary examples, however.  These include
the actively flaring stars LHS 2243 \citep[M8]{giz00} and LHS 2065 \citep[M9]{mrt01,scm02}, which have 3 cm
luminosity limits $\log{L_{\nu,q}/L_{bol}} < -17.0$ and $-17.5$, respectively \citep[B02]{kri99}.
Compared to predicted values from Eqn.\ 5,
$-17.2$ and $-17.1$, these upper limits are below, but still fall within 0.5 dex of,
this possible radio emission/spectral type trend.  Interestingly, both of these
undetected flare stars are slowly
rotating, with $v\sin{i} < 12$ km s$^{-1}$ \citep{moh03}.
Future monitoring observations of radio detected and undetected sources
will be needed to explore any correlation between quiescent radio emission and optical flaring.

\subsection{Violations of the G\"{u}del-Benz Relations}

As discussed in $\S$ 1, one
of the interesting revelations of the quiescent and flaring radio
emission from LP 944-20 and other late-type M and L dwarfs was the gross violation of the
radio/X-ray GB relations.
The sources detected in our sample also violate these relations.
The ROSAT X-ray detection of LHS 3003, assuming it to be
quiescent emission, implies $L_{\nu,q} \approx 10^{-15.5}L_X \approx 6{\times}10^{10}$
erg s$^{-1}$ Hz$^{-1}$, about 200 times fainter than measured here.
Upper limits on the X-ray emission from DENIS 1048$-$3956
predict radio fluxes $\sim$60 times fainter than our detection.
These deviations are not as extreme as those reported by B02 for the M9 dwarfs
LP 944-20 and BRI 0021$-$0214 (3-4
orders of magnitude), suggesting that the shift away
from the radio/X-ray empirical trend occurs gradually around spectral type M7.

The reason for this deviation is likely related to the
emission trends diagrammed in Figure 5.  If X-ray and optical emission are
correlated, as suggested by the relation $L_X \sim L_{H\alpha}$ typical
for M stars \citep{rei95,fle03}, the divergence of H$\alpha$ and radio
emission trends at and beyond spectral types M7 would be consistent with violations
of the GB relations.
The implication is that high-energy electrons in magnetic fields are present
around ultracool dwarfs, but that coronal and chromospheric plasma heating is somehow
suppressed or attenuated.  While electron densities still appear to be relatively
high in the radio-emitting coronal region (although this depends on the adopted emission mechanism),
chromospheric densities might be reduced in accordance
with the increasingly neutral photospheres of these objects.
This would explain the divergence of H$\alpha$ and radio emission trends,
but not necessarily the apparent divergence between X-ray and radio emission.
The latter may require a substantial decrease in the temperature
of coronal plasmas around ultracool stars and brown dwarfs,
perhaps due to a reduction
in the energetics of whatever nonthermal processes supply the radio-emitting region
with ionized material.

\section{Where do the Coronal Plasmas Come From?}

The presence of magnetic fields around late-type M, L, and even
cooler brown dwarfs is in itself not surprising, as large-scale
fields are generated by the Solar giant planets despite having
T$_{eff} \lesssim 130$ K.  Indeed, the dynamo mechanism for
a gas giant such as Jupiter, driven by
convective motions in the fluid metallic Hydrogen interior \citep{ste03}, may be a good analogy
for the dynamo mechanism employed by ultracool stars and brown dwarfs.
A more intriguing mystery
is the origin and retention of coronal and chromospheric plasmas
in the context of increasingly neutral photospheres.
Electron precipitation is the dominant loss mechanism of the coronal plasma,
occurring over timescales (${\tau}_e$) of order minutes \citep{lin83,kun87}.  The
presence of quiescent radio emission necessitates a constant
replenishment of coronal plasma, but
how is this plasma supplied?
The most commonly prescribed source is microflaring:
short, rapid, bursting emission that continually
accelerates electrons to coronal energies and may in fact comprise the
observed quiescent flux \citep{gud02}.  Evidence of substantial variability or polarization
in quiescent radio emission would provide support for this interpretation.  While our
detections are too close to the sensitivity limit to usefully
test this prediction, the
detection of multiple flaring events on LP 944-20 (B01) and highly variable
emission from the L3.5 2MASS 0036+1821 (B02) are certainly supportive.
Furthermore, the possible correlation between radio emission and optical flaring
suggested above could provide a means of moving ionized material into the upper
atmosphere.

External sources for coronal plasma should also be considered. One possibility is
accretion from the interstellar medium (ISM), which can be expressed as
\begin{equation}
(\frac{dN_e}{dt})_{in} \sim {\epsilon}n_{ISM}{\pi}{\cal L}^2V,
\end{equation}
where $N_e \sim n_e{\cal L}^3$ is the total number of electrons in the
corona, $n_{ISM} \approx 0.07$ cm$^{-3}$ is the density of the ISM \citep{par84},
$V \sim 30$ km s$^{-1}$ is the relative dwarf/ISM velocity (roughly the
typical space velocity of a late-type disk dwarf; Gizis et al.\ 2000),
and $\epsilon$ is a numerical factor encompassing the fraction of ISM material acquired and ionized
by the passing dwarf.  A sustained coronal plasma requires
$dN_e/dt = -(dN_e/dt)_{precip} + (dN_e/dt)_{in} = 0$, where
$(dN_e/dt)_{precip} \sim N_e/{\tau}_e$.
Even assuming\footnote{Perfect acquisition of ISM material is
unlikely for a number of reasons, including the presence of stellar winds and shielding
by the magnetosphere.  This assumption is therefore overoptimistic.}
$\epsilon$ = 1,
the condition of equilibrium results in a coronal
plasma density $n_e \sim 10^{-2}$ cm$^{-3}$, several orders of magnitude less than observed.
Hence, ISM accretion is not a viable method.

Another possibility is ongoing accretion from a circumstellar disk.  Assuming
the accretion of
hydrogen gas at a rate $\dot{\rm M}$,
\begin{equation}
(\frac{dN_e}{dt})_{in} \sim \frac{{\epsilon}\dot{\rm M}}{{\rm m}_p},
\end{equation}
where ${\rm m}_p = 1.7{\times}10^{-24}$ g is the proton mass.  For an accretion rate
of 10$^{-10}$ M$_{\sun}$ yr$^{-1}$, derived for 50 Myr M-type brown dwarfs in the R CrA
association \citep{bar04},
an equilibrium coronal density of $5{\times}10^{10}$ cm$^{-3}$ can be maintained for
$\epsilon$ = 0.1.
Hence, young stars and brown dwarfs could sustain coronal plasmas
through accretion alone.  On the other hand,
this mechanism may not be viable for older field dwarfs, as disk accretion
drops off dramatically for ages $\gtrsim$ 50 Myr \citep{bou97}.

Finally, plasma accretion could originate
from a close companion undergoing steady mass loss.
This mechanism is responsible for
Jupiter's auroral plasma, the majority
of which is supplied from the tidally stressed moon Io.
If planetary systems analogous to Jupiter's moon system exist around
very low-mass stars and brown dwarfs, accretion from those objects is a viable, albeit rare possibility
(due to tidal circularization and the special geometry of the Io-Europa-Ganymede system).
For Jupiter, the current flow between Io and
Jupiter also gives rise to
strong, variable decametric radio emission \citep{bur55,big64,gol69}, and
it is possible that coherent emission from
DENIS 1048$-$3956 is a high-frequency analogue of this interaction.
Searches for similar radio emission from systems with known extrasolar
planets have thus far turned up negative \citep{win86,bst00}, although
most of these systems contain early-type, relatively inactive primaries.
If the primary is a magnetically active late-type M star, the
likelihood of ECM emission at GHz frequencies may be increased.
We note that magnetic star-planet interactions have been suggested as a means of inducing
mass ejection \citep{ip04}, a possible explanation for the apparent
chromospheric mass motion observed in the
optical flare of DENIS 1048$-$3956 \citep{fuh04}.

These speculative hypotheses for the origin and retention
of coronal plasmas reflect both poor observational constraints
and limited modelling of the coronae of cool dwarf stars and brown dwarfs.
One point is certain, however; substantial and sustained
ionized material is present in the upper atmospheres of these objects
despite the observed trends in optical and X-ray emission and
theoretical expectations.  Future studies of the variability, physical
extent, and spectral characteristics of the radio emission may
help ascertain the nature of this hot coronal gas.

\section{Summary}

From a sample of seven late-type M and L dwarfs within 20 pc from the Sun,
we have detected two late-type M dwarfs, LHS 3003 and DENIS 1048$-$3956,
in quiescent emission at 4.80 GHz.  This emission indicates
that both magnetic fields ($B \gtrsim 10$ G)
and sustained coronal plasmas ($n_e \lesssim 10^{9}$ cm$^{-3}$) are present in
these sources, contrary to theoretical expectations.  Coupled with VLA detections
of ultracool dwarfs by B01 and B02, there is an apparent trend for radio emission to remain
constant or increasing over spectral types M5 to M9, and possibly into the L dwarf regime,
coincident with a rapid decline in optical
(H$\alpha$) and X-ray emission.  The deviation
in these activity trends explains gross violations of the GB relations, and
indicates a shift in the magnetic emission mechanisms of active stars and brown dwarfs around
spectral type M7/M8.

We also detected DENIS 1048$-$3956 in
two strong, rapid, and highly polarized flares at both 4.80 and 8.64 GHz, each 4-5
minutes in duration separated
by 10 minutes.  These flares have a coherent emission origin,
and their similarity to highly polarized bursts from other active M stars
suggests plasma or ECM emission
from a small (${\cal L} \lesssim 0.04$ R$_*$), high density ($n_e \sim 10^{11}$-$10^{12}$
cm$^{-3}$), highly magnetic ($B \sim 1$ kG) region.  The temporal proximity
of the flaring events suggests a large-scale frequency drift in the emission, possibly due to
motion or compression of the emitting region.  Both persistent and flaring radio
emissions make up a small ($L_R \sim 10^{-7}$-$10^{-6}L_{bol}$) but nontrivial
fraction of the total luminosity from this low-mass star.

To date, seven nearby ultracool dwarfs spanning spectral types M7 to L5 have been reported
in quiescent and flaring radio emission.  All of the detected sources lie within 12 pc
of the Sun, implying that a significant fraction of cool dwarfs overall
are in fact radio-emitters.  While the relative radio emission shows some indication of
increasing with spectral type, it remains unclear as to whether
there is any strict correlation
with rotation or the presence of optical emission, although rapidly
rotating and/or flaring sources are more often radio-emitters.
Clearly, a larger sample of well-characterized
objects is needed to explore these trends.  While the detection of radio emission
proves the presence of magnetic fields and coronal plasmas
on ultracool dwarfs, there remains substantial uncertainty as to the strength,
scale, stability, and origin of the fields and plasmas, which should be explored
with further observations and theoretical work.

\acknowledgments

A.\ J.\ B.\ and M.\ E.\ P.\ would like to thank members of the Australia Telescope
National Facility for their support and
hospitality during the observations, and in particular our duty astronomers
Jules Harnett and Jess O'Brien.  We also thank N.\ McClure-Griffiths for assisting in observations
during our May 2002 run.
We give special thanks to our anonymous referee, whose extensive comments allowed us
to greatly improve the original manuscript.
A.\ J.\ B.\ acknowledges useful discussions with T.\ Ayres, E.\ Berger, A.\ Burrows, M.\ Giampapa,
J.\ Liebert, J.\ Linsky, A.\ Melatos, J.\ Turner, \& R.\ Webster
during the preparation of this manuscript.
Support for this work was provided by NASA through Hubble Fellowship grants from the
Space Telescope Science Institute, which is operated by the Association of Universities
for Research in Astronomy, Incorporated, under NASA contract NAS5-26555;
and through the Spitzer Fellowship Program.
The Australia Telescope is funded by the Commonwealth of Australia for operation
as a National Facility managed by CSIRO.

Facilities: \facility{ATCA}.

\clearpage

\begin{figure}
\plottwo{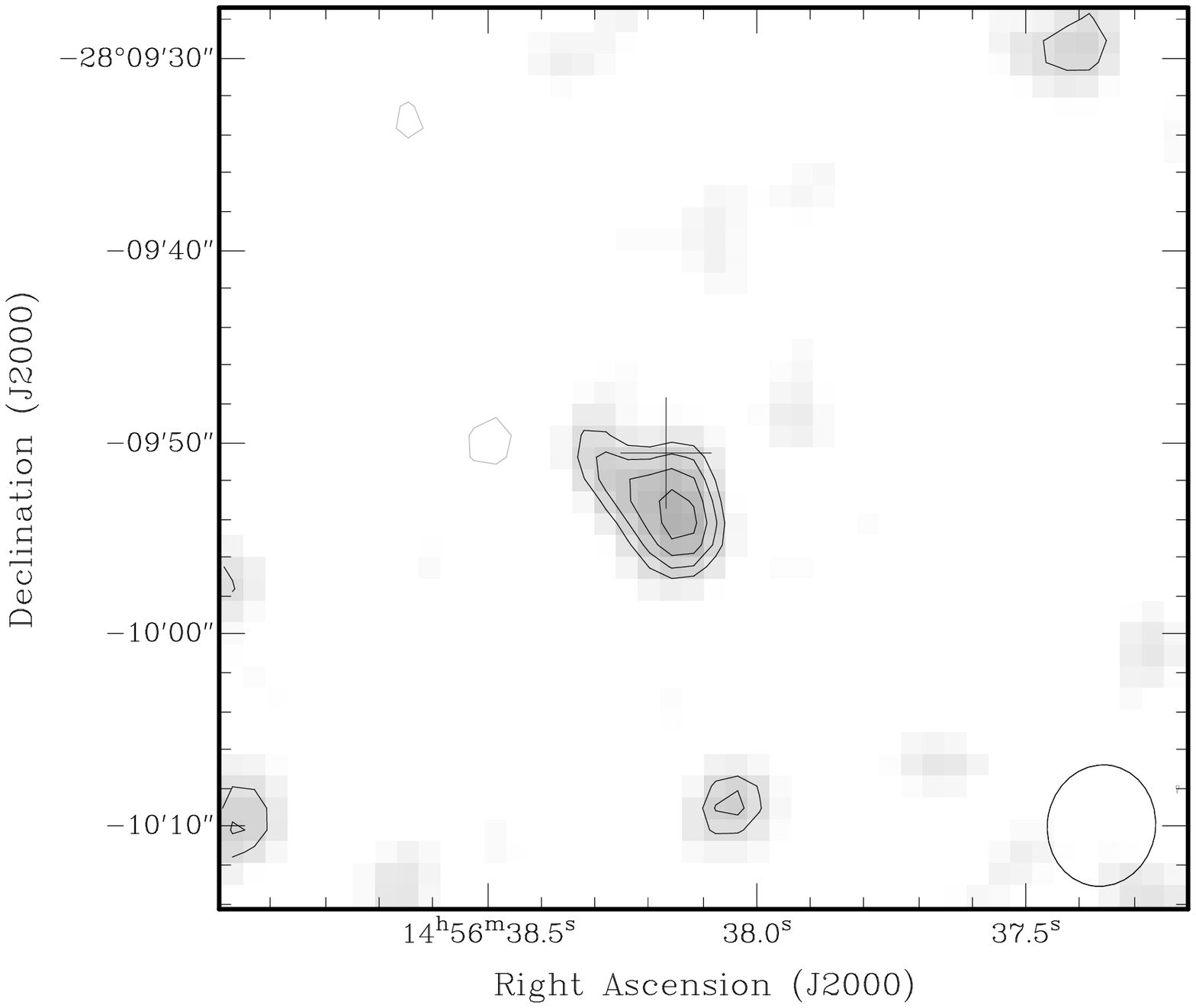}{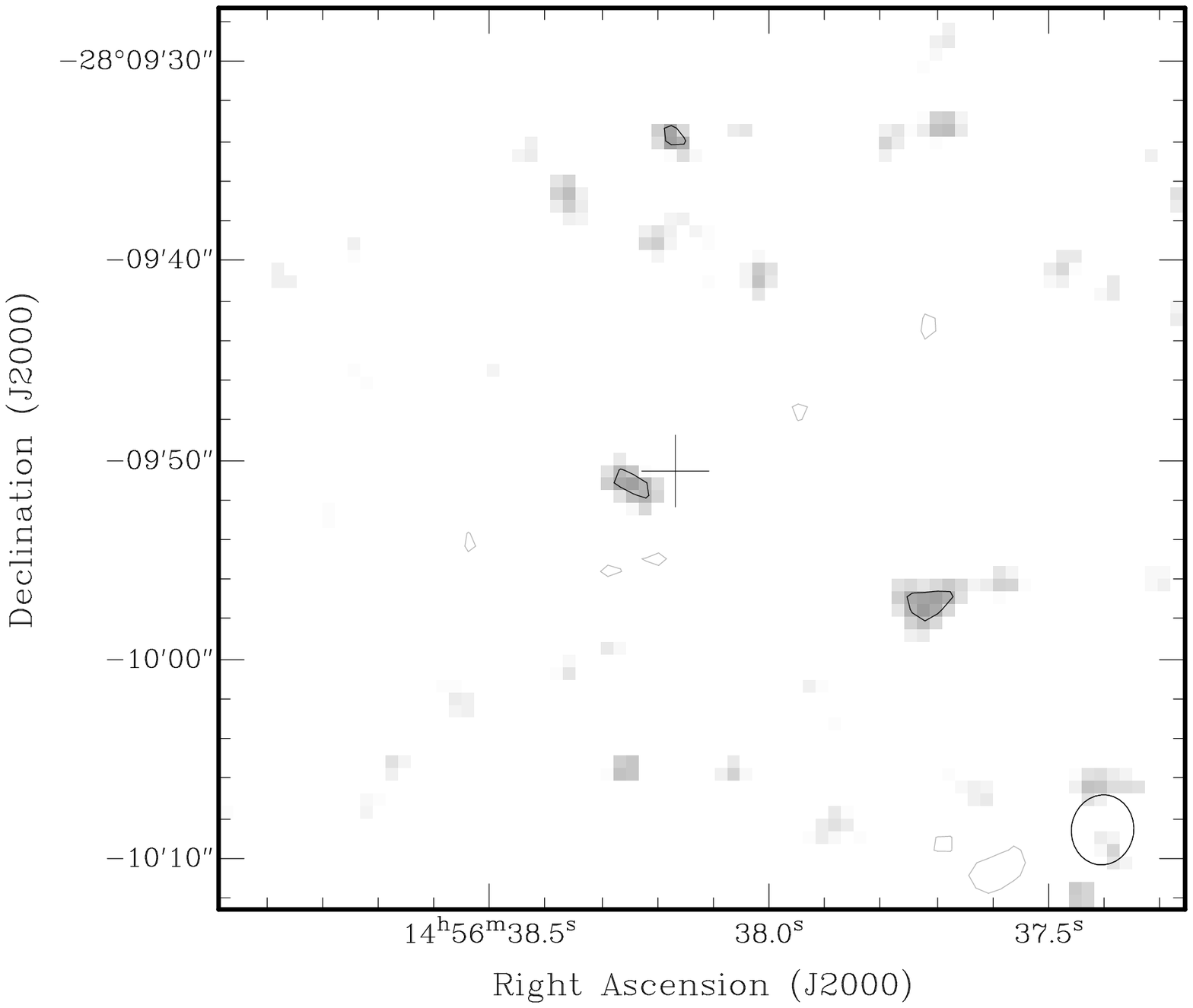}
\caption{Cleaned Stokes I images of the M7 dwarf LHS 3003 at 4.80 (left) and 8.64 GHz (right).
Images are roughly 45$\arcsec$ on a side oriented with north up
and east to the left. The beam shape for each frequency is shown in the bottom right corner.
Flux density contour lines of -0.1, 0.1, 0.125, 0.15, and 0.175 mJy beam$^{-1}$
are shown.  The expected location of LHS 3003 is indicated by a the large cross
at center. \label{fig1}}
\end{figure}

\begin{figure}
\plottwo{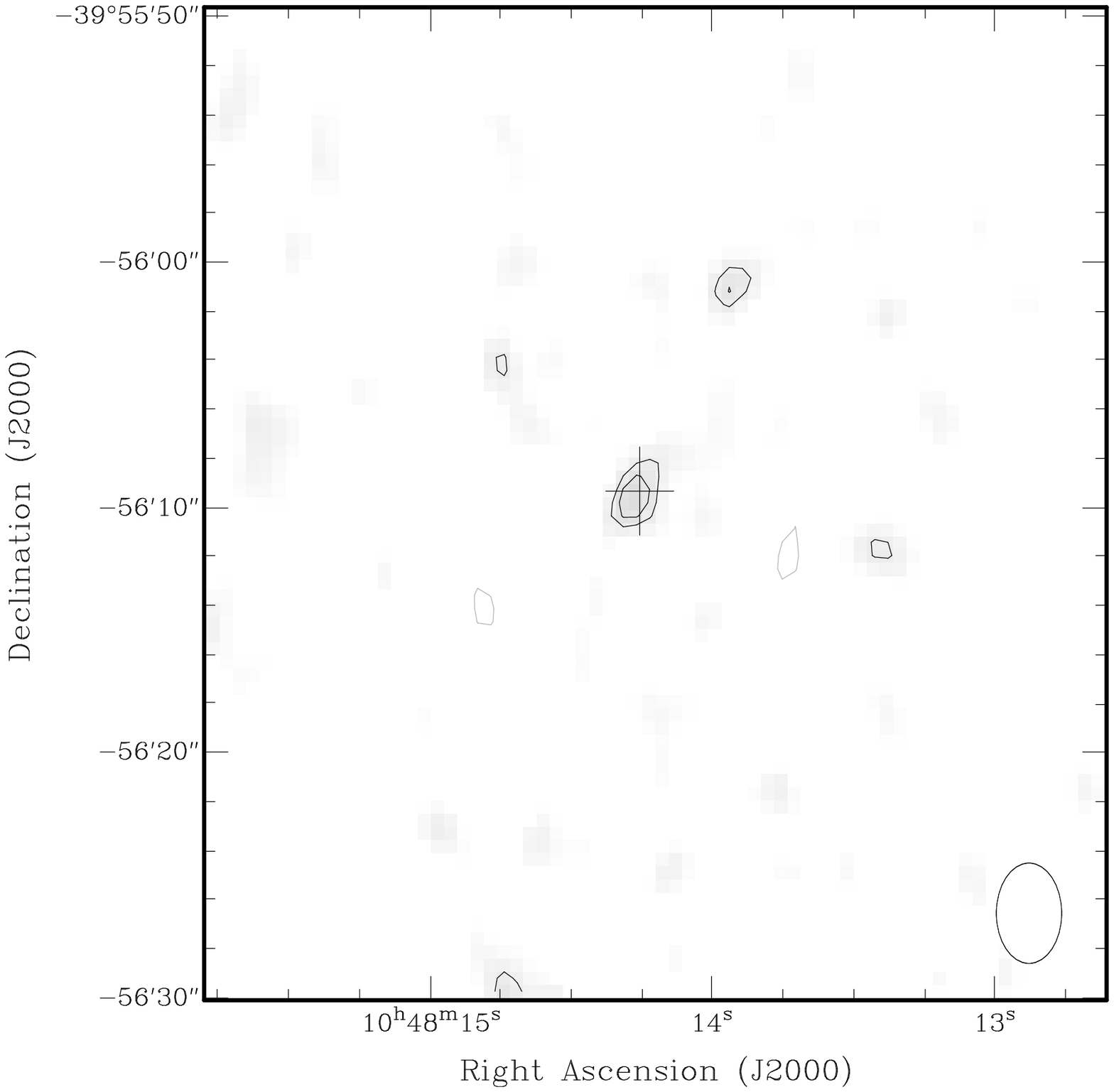}{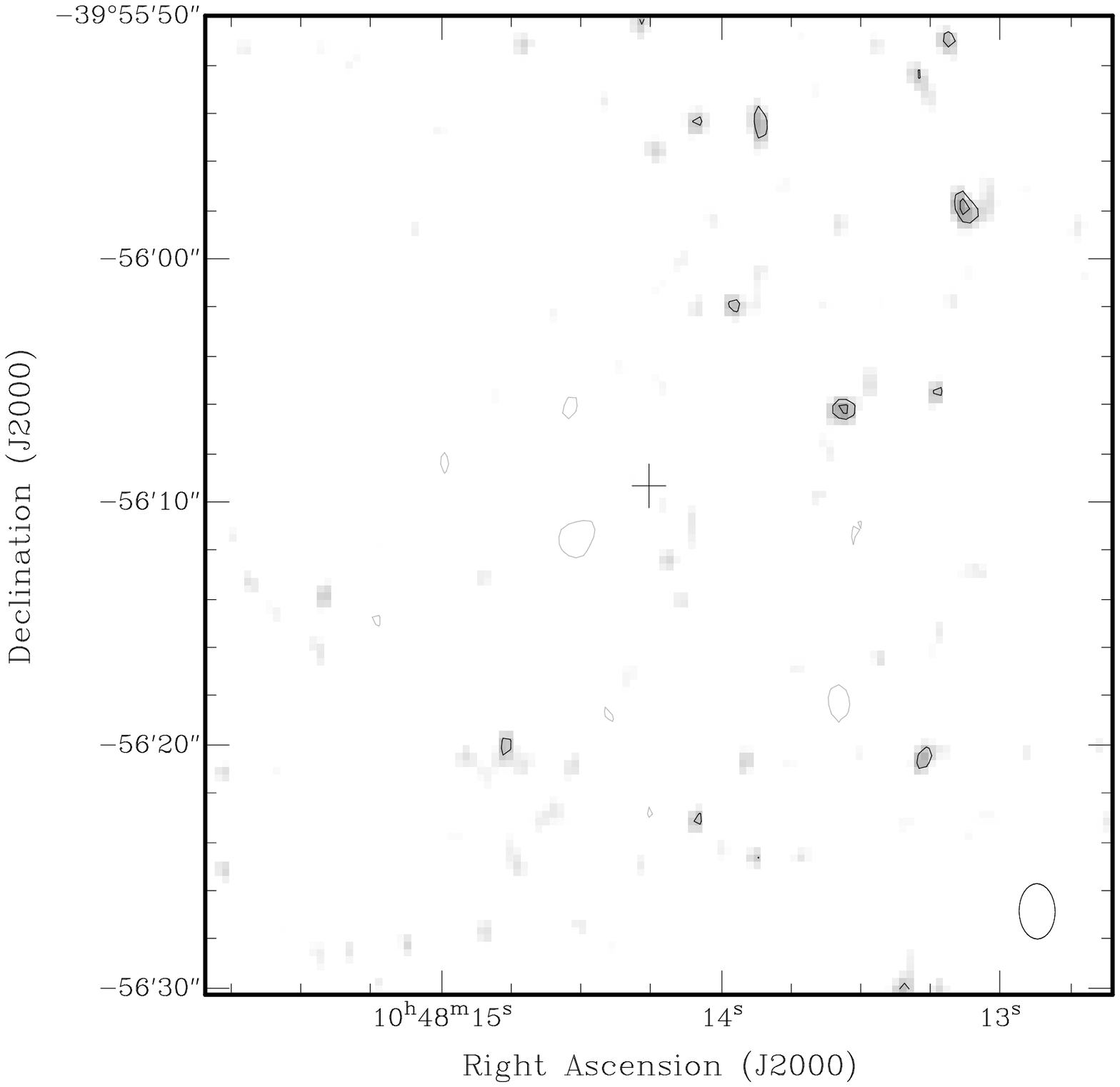}
\caption{Same as Figure 1 for the M8 DENIS 1048$-$3956. Images are 40$\arcsec$ on a side.
Visibility data during the observed flares ($\S$ 4) have been excluded from these images.
\label{fig2}}
\end{figure}

\begin{figure}
\plotone{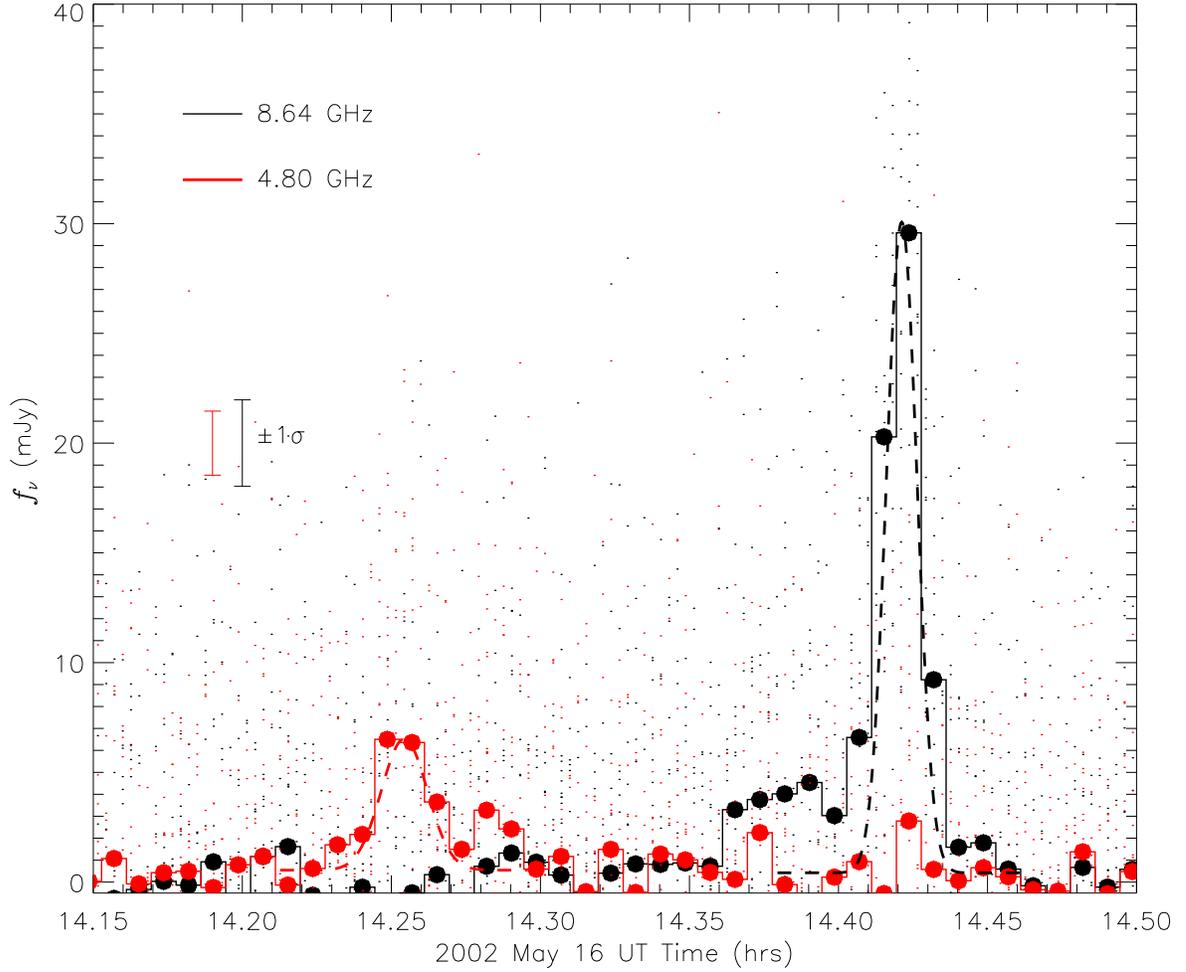}
\caption{Time series data for the DENIS 1048-3956 flaring emission
at 4.80 (red) and 8.64 GHz (black).  The individual flux-calibrated visibility
measurements are shown as small points; note the clear peak up in these data during
the 8.64 GHz flare.  Time-averaged
visibilities (all baselines, central ${\Delta}{\nu} = 72$ MHz,
and 30 sec time resolution) are indicated by
filled circles and histograms, and 1$\sigma$ error bars (based on the fluctuation
of the binned data outside the flare periods) are indicated.
The median background emission
has been subtracted from these data.
Gaussian fits to the unbinned visibility data during the flares at 14:15:15 (4.80 GHz)
and 14:25:16 UT (8.64 GHz)
are indicated by dashed lines.  \label{fig5}}
\end{figure}

\begin{figure}
\plottwo{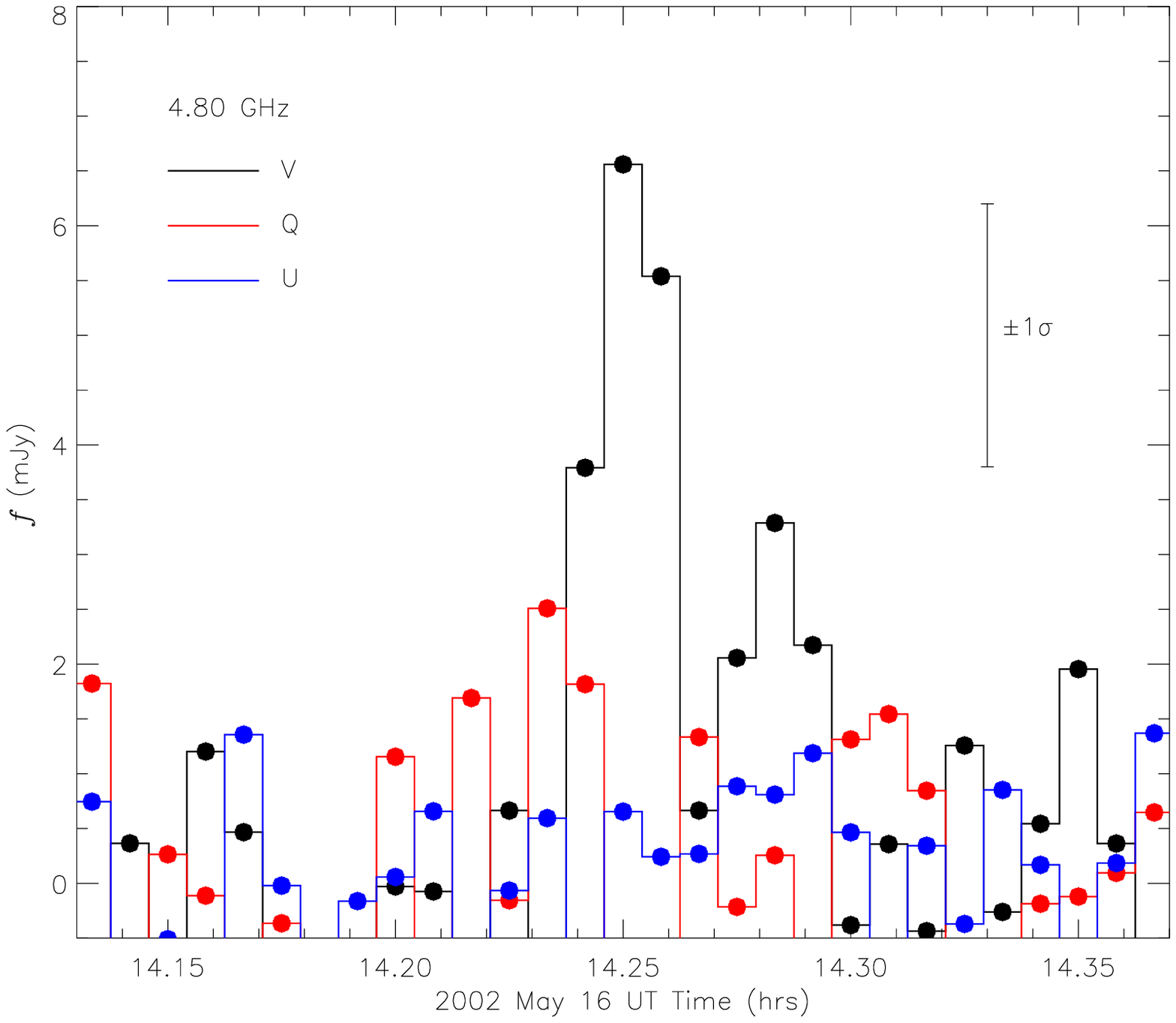}{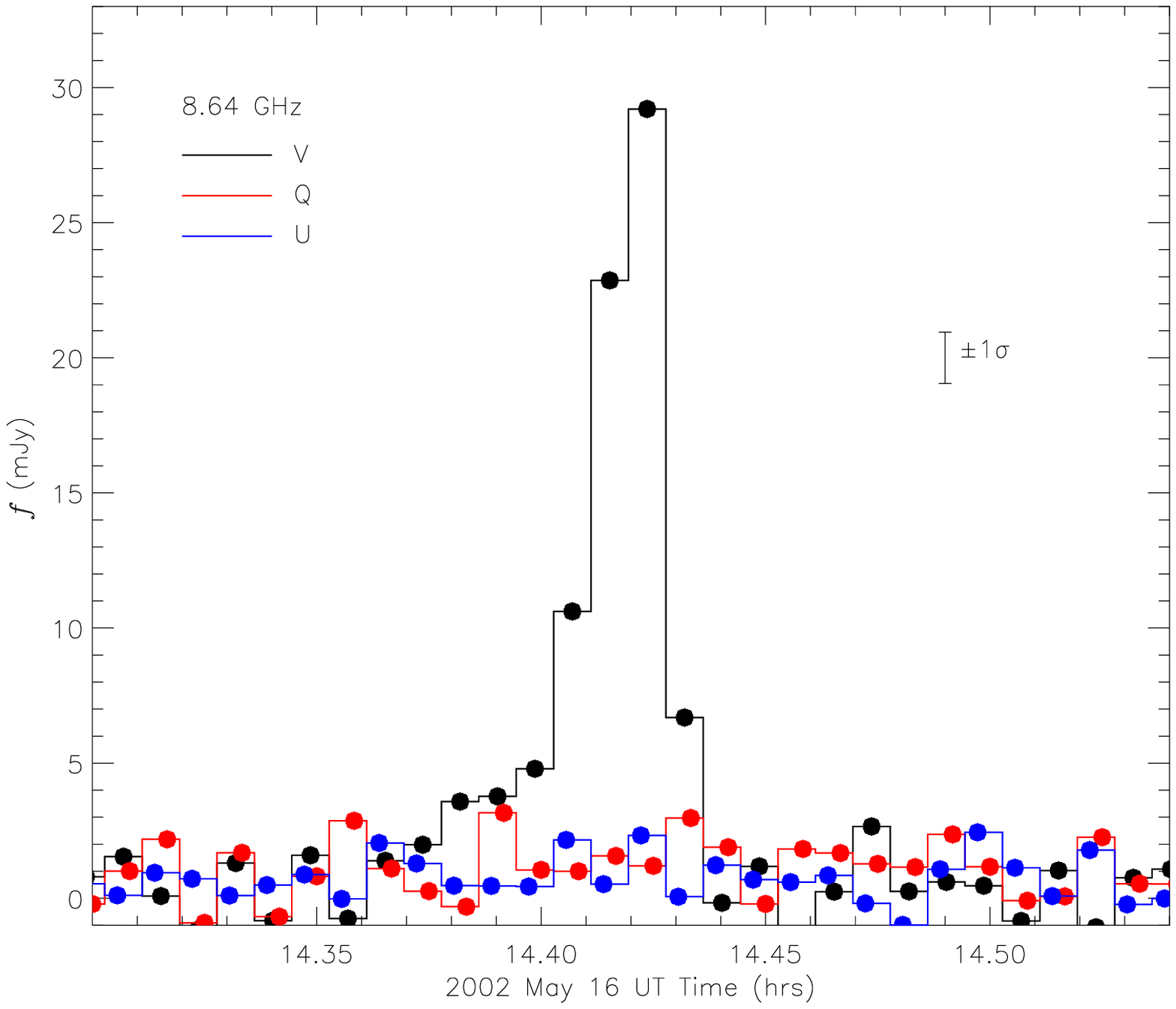}
\caption{Time series data for the 4.80 (left) and 8.60 GHz (right) flares in
Stokes V (black), Q (red), and U (blue)
polarizations.
Time-binned visibility data (30 s) for periods of 15 min about the flare peaks
are indicated by filled circles and histograms.
The median background/continuum emission
has been subtracted from these data.
Error bars (1$\sigma$, maximum for all three
polarizations) were computed as in Figure 3. \label{fig6}}
\end{figure}

\begin{figure}
\plotone{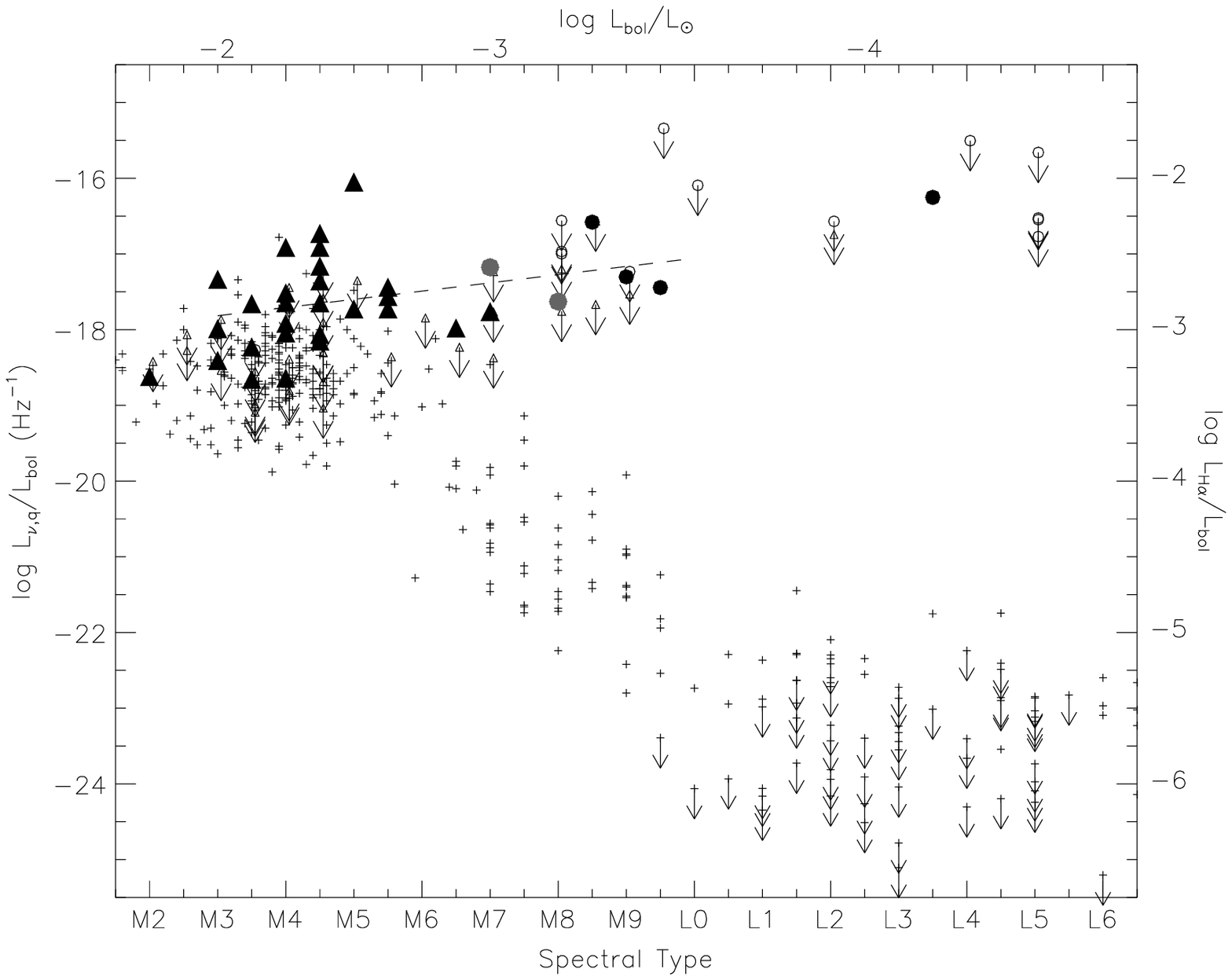}
\caption{Ratio of quiescent radio (left axis) and H$\alpha$ (right axis)
to bolometric luminosity versus spectral type for late-type main sequence stars.  M
dwarf radio data from \citet{lin83,whi89,gud93,kri99}; and \citet{let00} are indicated by triangles,
while ultracool (late-type M and L dwarf) data from B01, B02 (black), and this paper (grey)
are indicated by circles.  Open symbols with arrows indicate
upper limits.
H$\alpha$ data from \citet{haw96}; \citet{giz00}; and \citet{me02b} are shown as crosses, and
upper limits are indicated by arrows.
Bolometric luminosities for all sources were
derived from a seventh order polynomial fit to empirical values from the 8 pc sample
\citep{rei00a} and \citet{gol04} for spectral types M0 to L7;
standard deviation in the fit was 0.35 mag.
A linear fit to $\log{L_{\nu,q}/L_{bol}}$ for M3-M9 detected
radio sources (Eqn.\ 5) is indicated by the dashed line. \label{fig7}}
\end{figure}

\clearpage

\begin{deluxetable}{lcclcccccccl}
\tabletypesize{\scriptsize}
\rotate
\tablecaption{Late-type M and L Dwarf Targets. \label{tbl1}}
\tablewidth{0pt}
\tablehead{
 & \multicolumn{2}{c}{Coordinates\tablenotemark{a}} &
 & & & & &  &  \multicolumn{2}{c}{H$\alpha$ Emission} &  \\
\cline{2-3} \cline{10-11}
\colhead{Name} &
\colhead{$\alpha_{J2000}$} &
\colhead{$\delta_{J2000}$} &
\colhead{SpT} &
\colhead{$d$} &
\colhead{T$_{eff}$} &
\colhead{$\log{L_{bol}}$} &
\colhead{$\log{L_X}$} &
\colhead{$v\sin{i}$} &
\colhead{Quies.} &
\colhead{Flare} &
\colhead{References} \\
\colhead{} &
\colhead{} &
\colhead{} &
\colhead{} &
\colhead{(pc)} &
\colhead{(K)} &
\colhead{(erg s$^{-1}$)} &
\colhead{(erg s$^{-1}$)} &
\colhead{(km s$^{-1}$)} &
\colhead{} &
\colhead{} &
\colhead{} \\
\colhead{(1)} &
\colhead{(2)} &
\colhead{(3)} &
\colhead{(4)} &
\colhead{(5)} &
\colhead{(6)} &
\colhead{(7)} &
\colhead{(8)} &
\colhead{(9)} &
\colhead{(10)} &
\colhead{(11)} &
\colhead{(12)} \\
}
\startdata
LHS 102A & 00$^h$04$^m$37$\fs$06 & $-$40$\degr$44$\arcmin$07$\farcs$7 & M3.5 & 9.55$\pm$0.10 & 3200 & 31.3 & $<$ 27.5 & ... & No & ...  & 1,2,3,4 \\
LHS 102B & 00$^h$04$^m$35$\fs$07 & $-$40$\degr$44$\arcmin$11$\farcs$5 & L5 & 9.55$\pm$0.10 & 1900 & 29.6 & $<$ 27.5 & 32.5$\pm$2.5 & Yes & ...  & 2,4,5,6 \\
SSSPM 0109$-$5100 & 01$^h$09$^m$01$\fs$53 & $-$51$\degr$00$\arcmin$49$\farcs$7 & L2 & $\sim$ 10 & 2100 & 29.7 & $<$ 27.5 & ... & ... & ...   & 7,8,9 \\
2MASS 0835$-$0819 & 08$^h$35$^m$42$\fs$56 & $-$08$\degr$19$\arcmin$23$\farcs$7 & L5 & $\sim$ 8 & 1700 & 29.5 & $<$ 27.4 & ... & No & ...  & 8,9,10 \\
DENIS 1048$-$3956 & 10$^h$48$^m$14$\fs$26 & $-$39$\degr$56$\arcmin$09$\farcs$3 & M8 & 4.6$\pm$0.3 & 2500 & 30.2 & $<$ 26.3 & 25$\pm$2 & Yes & Yes & 9,11,12,13,14,15 \\
2MASS 1139$-$3159\tablenotemark{b} & 11$^h$39$^m$51$\fs$11 & $-$31$\degr$59$\arcmin$21$\farcs$1 & M8 & $\sim$ 20 & 2500 & 30.2 & $<$ 28.2 & ... & Yes & ...  & 8,9,12 \\
LHS 3003 & 14$^h$56$^m$38$\fs$17 & $-$28$\degr$09$\arcmin$50$\farcs$5 & M7 & 6.56$\pm$0.15 & 2600 & 30.3 & 26.3 & 8.0$\pm$2.5 & Yes & Yes  & 4,16,17,18,19,20 \\
2MASS 1534$-$1418 & 15$^h$34$^m$57$\fs$04 & $-$14$\degr$18$\arcmin$48$\farcs$6 & M8 & $\sim$ 11 & 2500 & 30.2 & $<$ 27.6 & ... & Yes & ...  & 7,8,12 \\
\enddata
\tablenotetext{a}{J2000 coordinates from 2MASS (epoch 1998.5--1999.5)
updated to the observation epoch (except for 2MASS 0835$-$0819 and 2MASS 1534$-$1418) using proper motion
measurements from \citet{tin96,del01,giz02}; and \citet{sch02}.}
\tablenotetext{b}{TW Hyd candidate \citep{giz02}.}
\tablerefs{(1) \citet{luy79a}; (2) \citet{van95}; (3) \citet{haw96}; (4) \citet{leg02};
(5) \citet{gld99}; (6) \citet{bas00}; (7) \citet{lod02};
(8) distance estimated using $M_J$/spectral type relation of \citet{cru03};
(9) T$_{eff}$ and L estimated using spectral type relations of \citet{gol04};
(10) \citet{cru03}; (11) \citet{del01}; (12) \citet{giz02}; (13) \citet{neu02}; (14) \citet{fuh04};
(15) \citet{scm04}; (16) \citet{bes91}; (17) \citet{ian95}; (18) \citet{sch95}; (19) \citet{moh03};
(20) \citet{rui90}}
\end{deluxetable}

\begin{deluxetable}{lllcccl}
\tabletypesize{\scriptsize}
\tablecaption{Log of Observations. \label{tbl2}}
\tablewidth{0pt}
\tablehead{
 & \multicolumn{2}{c}{UT Time} &  & \multicolumn{2}{c}{Beam Size}  & \\
\cline{2-3}\cline{5-6}
\colhead{Object} &
\colhead{Start} &
\colhead{Finish} &
\colhead{$t_{track}$ (hr)} &
\colhead{4.80 GHz} &
\colhead{8.64 GHz} &
\colhead{Secondary Cal} \\
\colhead{(1)} &
\colhead{(2)} &
\colhead{(3)} &
\colhead{(4)} &
\colhead{(5)} &
\colhead{(6)} &
\colhead{(7)} \\
}
\startdata
LHS 102AB & 2002 May 16 16:15 & May 17 03:29 & 11.2  & 4$\farcs$2$\times$2$\farcs$6 & 2$\farcs$4$\times$1$\farcs$4 & PKS B0008-421 \\
SSSPM 0109-5100 & 2002 Dec 02 06:05 & Dec 02 17:45 & 11.7 & 3$\farcs$5$\times$2$\farcs$8 & 1$\farcs$9$\times$1$\farcs$6 & PKS B0047-579 \\
2MASS 0835$-$0819 & 2002 Nov 29 13:16 & Nov 30 23:02 & 9.8 & 25$\farcs$7$\times$2$\farcs$2 & 14$\farcs$3$\times$1$\farcs$2 & PKS B0859-140 \\
DENIS 1048$-$3956 & 2002 May 16 03:11 & May 16 15:32 & 12.4 & 4$\farcs$1$\times$2$\farcs$7 & 2$\farcs$3$\times$1$\farcs$5 &  PKS B1104-445 \\
2MASS 1139$-$3159 & 2002 Nov 30 14:47 & Dec 01 01:39 & 10.9 &  4$\farcs$9$\times$2$\farcs$6 & 2$\farcs$7$\times$1$\farcs$5 & PKS B1144-379 \\
LHS 3003 & 2002 May 17 07:12 & May 17 18:09 & 11.0 & 6$\farcs$3$\times$2$\farcs$7 & 3$\farcs$5$\times$1$\farcs$5 & PKS B1514-241 \\
2MASS 1534$-$1418 & 2002 Dec 01 19:24 & Dec 02 05:28 & 10.1 & 13$\farcs$0$\times$2$\farcs$5 & 7$\farcs$2$\times$1$\farcs$4 & PKS B1504-166 \\
\enddata
\end{deluxetable}

\begin{deluxetable}{lcccccc}
\tabletypesize{\scriptsize}
\tablecaption{Quiescent Emission. \label{tbl3}}
\tablewidth{0pt}
\tablehead{
\colhead{Object} &
\colhead{$f_{4.80}$ (mJy)} &
\colhead{$f_{8.64}$ (mJy)} &
\colhead{$\alpha$} &
\colhead{$\log L_{\nu,q}$ (erg s$^{-1}$ Hz$^{-1}$)} &
\colhead{$\log L_{R}/L_{bol}$\tablenotemark{a}} &
\colhead{T$_B$ (K)\tablenotemark{b}} \\
\colhead{(1)} &
\colhead{(2)} &
\colhead{(3)} &
\colhead{(4)} &
\colhead{(5)} &
\colhead{(6)} \\
}
\startdata
LHS 102A & $<$ 0.09 & $<$ 0.11 & ... & $<$ 13.0 & $< -8.4$ & $< 2{\times}10^8$ \\
LHS 102B & $<$ 0.09 & $<$ 0.11 & ... & $<$ 13.0 & $< -6.7$ & $< 2{\times}10^8$ \\
SSSPM 0109$-$5100 & $<$ 0.11 & $<$ 0.11 & ... & $<$ 13.1 & $< -6.7$&  $< 7{\times}10^7$ \\
2MASS 0835$-$0819 & $<$ 0.12 & $<$ 0.12 & ... & $<$ 13.0 & $< -6.6$ & $< 5{\times}10^7$ \\
DENIS 1048$-$3956 & 0.14$\pm$0.04 & $<$ 0.11 & $< -0.4$ & 12.6 & $-7.7$ & $(3-6){\times}10^7$ \\
2MASS 1139$-$3159 & $<$ 0.12 & $<$ 0.10 & ... & $<$ 13.7 & $< -6.5$ & $< 3{\times}10^8$ \\
LHS 3003 & 0.27$\pm$0.04 & $<$ 0.12 & $<$ $-1.2$ & 13.1 & $-7.3$ & $(1-3){\times}10^8$ \\
2MASS 1534$-$1418 & $<$ 0.11 & $<$ 0.11 & ... & $<$ 13.2 & $< -7.1$ & $<$ $9{\times}10^7$ \\
\enddata
\tablenotetext{a}{Assuming a spectral energy distribution peaked at ${\nu}_{pk} = 4.80$ GHz,
with $f_{\nu} \propto {\nu}^{2.5}$ below ${\nu}_{pk}$ and
$f_{\nu} \propto {\nu}^{-1.5}$ above ${\nu}_{pk}$ over the range ${\nu}_{pk}/10 < \nu < 10{\nu}_{pk}$; see $\S$ 3.2.}
\tablenotetext{b}{Assuming source dimension ${\cal L} \sim$ (2-3)$\times$R$_*$, where
R$_*$ $\approx$ 0.1R$_{\sun}$ $\approx$ 1 R$_{Jup}$ $= 7{\times}10^9$ cm
for late-type M and L dwarfs.
For the M3.5 V LHS 102A, we assume R$_*$ $\approx$ 0.2R$_{\sun}$ based
on interferometric radius measurements of similarly-typed M dwarfs by \citet{lan01} and \citet{seg03}.}
\end{deluxetable}

\begin{deluxetable}{lcc}
\tabletypesize{\scriptsize}
\tablecaption{Flaring Emission from DENIS 1048$-$3956 on 2002 May 16 (UT). \label{tbl4}}
\tablewidth{0pt}
\tablehead{
\colhead{} &
\colhead{4.80 GHz} &
\colhead{8.64 GHz} \\
}
\startdata
$t_{pk}$ (UT)\tablenotemark{a} & 14:15:15$\pm$6 s & 14:25:16$\pm$1 s \\
$f_{\nu,f}$ (mJy)\tablenotemark{a} & 6.0$\pm$0.8 & 29.6$\pm$1.0 \\
$\Pi_Q$ (\%)\tablenotemark{b} & $<$18 & $<$3 \\
$\Pi_U$ (\%)\tablenotemark{b} & $<$20 & $<$3 \\
$\Pi_V$ (\%)\tablenotemark{b} & $\sim$100 & $\sim$100 \\
T$_B({\cal L}/R_*)^{-2}$ (K)\tablenotemark{c} & $(1.1{\pm}0.2){\times}10^{10}$ & $(1.7{\pm}0.2){\times}10^{10}$ \\
$\log L_{\nu,f}$ (erg s$^{-1}$ Hz$^{-1}$) & 14.2 & 14.9 \\
\enddata
\tablenotetext{a}{Emission peak time and flux density based on Gaussian fits to the
unbinned times series data (Figure 3).}
\tablenotetext{b}{Polarizations at flare peak; upper limits for $\Pi_Q$ and $\Pi_U$
are estimated from 1$\sigma$ uncertainties in the time series data.}
\tablenotetext{c}{We assume R$_*$ $\approx$
R$_{Jup}$ $= 7{\times}10^9$ cm.  For ${\cal L} \lesssim 0.04$R$_*$,
$T_B \gtrsim 10^{13}$ K (see $\S$ 4.2).}
\end{deluxetable}


\begin{thebibliography}{}

\bibitem[Altshuler(1986)]{alt86} Altshuler, D.\ R. 1986, \aaps, 65, 267

\bibitem[Bailer-Jones(2004)]{bai04} Bailer-Jones, C.\ A.\ L. 2004, \aap, 419, 703

\bibitem[Barrado y Navascu{\'{e}}s, Mohanty, \& Jayawardhana(2004)]{bar04}Barrado y
Navascu{\'{e}}s, D., Mohanty, S., \& Jayawardhana, R. 2004, \apj, 604, 284

\bibitem[Basri \& Marcy(1995)]{bas95}Basri, G., \& Marcy, G.\ W. 1995, \aj, 109, 762

\bibitem[Basri et al.(2000)]{bas00}Basri, G., Mohanty, S., Allard, F.,
Hauschildt, P.\ H., Delfosse, X., Mart{\'{\i}}n, E.\ L., Forveille, T.,
\& Goldman, B. 2000, \apj, 538, 363

\bibitem[Bastian(1990)]{bas90}Bastian, T.\ S. 1990, SoPh, 130, 265

\bibitem[Bastian, Dulk, \& Leblanc(2000)]{bst00}Bastian, T.\ S., Dulk, G.\ A.,
\& LeBlanc, Y. 2000, \apj, 545, 1058

\bibitem[Benz, Alef, \& G\"{u}del(1995)]{ben95}Benz, A.\ O., Alef, W.,
\& G\"{u}del, M. 1995, \aap, 298, 187

\bibitem[Benz, Conway, \& G\"{u}del(1998)]{ben98}Benz, A.\ O., Conway, J.,
\& G\"{u}del, M. 1998, \aap, 331, 596

\bibitem[Benz \& G\"{u}del(1994)]{ben94}Benz, A.\ O.,
\& G\"{u}del, M. 1994, \aap, 285, 621

\bibitem[Berger(2002)]{brg02}Berger, E. 2002, \apj, 572, 503 (B02)

\bibitem[Berger et al.(2001)]{brg01}Berger, E., et al. Nature, 410, 338 (B01)

\bibitem[Bessell(1991)]{bes91}Bessell, M.\ S. 1991, \aj, 101, 662

\bibitem[Bigg(1964)]{big64}Bigg, E.\ K. 1964, Nature, 203, 1088

\bibitem[Bingham, Cairns, \& Kellett(2001)]{bin01}Bingham, R., Cairns, R.\ A., \&
Kellett, B.\ J. 2001, \aap, 370, 1000

\bibitem[Bouvier, Forestini, \& Allain(1997)]{bou97} Bouvier, J., Forestini, M., \& Allain, S.
1997, \aap, 326, 1023

\bibitem[Burgasser et al.(2002a)]{me02b}Burgasser, A.\ J., Liebert, J.,
Kirkpatrick, J.\ D., \& Gizis, J.\ E. 2002a, \aj, 123, 2744

\bibitem[Burgasser et al.(2000)]{me00b}Burgasser, A.\ J., Kirkpatrick, J.\ D.,
Reid, I.\ N., Liebert, J., Gizis, J.\ E., \& Brown, M.\ E. 2000, \aj, 120,
473

\bibitem[Burgasser et al.(2002b)]{me02a} Burgasser, A.\ J., et al. 2002b, \apj, 564, 421

\bibitem[Burke \& Franklin(1955)]{bur55} Burke, B.\ F., \& Franklin, K.\ L. 1955, JGR, 60, 213

\bibitem[Burrows et al.(2001)]{bur01}Burrows, A., Hubbard, W.\ B., Lunine,
J.\ I., \& Liebert, J. 2001, Rev.\ of Modern Physics, 73, 719

\bibitem[Canfield et al.(1990)]{can90}Canfield, R.\ C., Metcalf, T.\ R., Zarro, D.\ M.,
\& Lemen, J.\ R. 1990, \apj, 348, 333

\bibitem[Chabrier \& Baraffe(1997)]{cha97}Chabrier, G., \& Baraffe, I.
1997, \aap, 327, 1039

\bibitem[Ciliegi et al.(2003)]{cil03}Ciliegi, P., Zamorani, G., Hasinger, G., Lehmann, I.,
Szokoly, G., \& Wilson, G. 2004, \aap, 398, 901


\bibitem[Condon et al.(1998)]{con98}Condon, J.\ J., Cotton, W.\ D., Greisen, E.\ W.,
Yin, Q.\ F., Perley, R.\ A., Taylor, G.\ B., \& Broderick, J.\ J. 1998, \aj, 115, 1693

\bibitem[Cruz et al.(2003)]{cru03} Cruz, K.\ L., Reid, I.\ N., Liebert, J., Kirkpatrick, J.\ D.,
\& Lowrance, P.\ J. 2003, AJ, 126, 2421

\bibitem[Cutri et al.(2003)]{cut03}Cutri, R.\ M., et al. 2003, Explanatory Supplement
to the 2MASS All Sky Data Release,
\url{http://www.ipac.caltech.edu/2mass/releases/allsky/doc/explsup.html}

\bibitem[Dahn et al.(2002)]{dah02}Dahn, C.\ C., et al. 2002, \aj,
124, 1170

\bibitem[Deacon \& Hambly(2001)]{dea01}Deacon, N.\ R., \& Hambly, N.\ C. 2001, \aap, 380, 148

\bibitem[de la Reza et al.(1989)]{del89} de la Reza, R., Torres, C.\ A.\ O., Quast, G.,
Castilho, B.\ V., \& Vieira, G.\ L. 1989, \apj, 343, L61

\bibitem[Delfosse et al.(2001)]{del01}Delfosse, X., et al. 2001, \aap,
336, L13

\bibitem[Donnelly, Partridge, \& Windhorst(1987)]{don87} Donnelly, R.\ H., Partridge, R.\ B.,
\& Windhorst, R.\ A. 1987, \apj, 321, 94

\bibitem[Dulk(1985)]{dul85}Dulk, G.\ A. 1985, \araa, 23, 169

\bibitem[Dulk \& Marsh(1982)]{dul82}Dulk, G.\ A., \& Marsh, K.\ A. 1982, \apj, 259, 350

\bibitem[Durney, De Young, \& Roxburgh(1993)]{dur93}Durney, B.\ R.,
De Young, D.\ S., \& Roxburgh, I.\ W. 1993, SoPh,
145, 207

\bibitem[Epchtein et al.(1997)]{epc97}Epchtein, N., et al. 1997,
The Messenger, 87, 27

\bibitem[Feigelson et al.(2002)]{fei02}Fiegelson, E.\ D., Broos, P., Gaffney, J.\ A., III,
Garmire, G., Hillenbrand, L.\ A., Pravdo, S.\ H., Townsley, L., \& Tsuboi, Y.
2002, \apj, 574, 258

\bibitem[Fleming, Giampapa \& Garza(2003)]{fle03}Fleming, T.\ A.,
Giampapa, M.\ S., \& Garza, D. 2003, \apj, 594, 982

\bibitem[Fleming, Giampapa \& Schmitt(2000)]{fle00}Fleming, T.\ A.,
Giampapa, M.\ S., \& Schmitt, J.\ H.\ M.\ M. 2000, \apj, 533, 372


\bibitem[Fomalont et al.(1991)]{fom91}Fomalont, E.\ B., Windhorst, R.\ A., Kristian, J.\ A.,
\& Kellermann, K.\ I. 1991, \apj, 475, L5

\bibitem[Fuhrmeister \& Schmitt(2004)]{fuh04}Fuhrmeister, B., \& Schmitt, J.\ H.\ M.\ M.
2004, \aap, 420, 1079


\bibitem[Gelino et al.(2002)]{gel02} Gelino, C.\ R., Marley, M.\
S., Holtzman, J.\ A., Ackerman, A.\ S., \& Lodders, K. 2002, \apj,
577, 433

\bibitem[Giampapa et al.(1996)]{gia96} Giampapa, M.\ S., Rosner, R., Kashyap, V.,
Fleming, T.\ A., Schmitt, J.\ H.\ M.\ M., \& Bookbinder, J.\ A. 1996, \apj, 463, 707

\bibitem[Gizis(2002)]{giz02}Gizis, J.\ E. 2002, \apj, 575, 484

\bibitem[Gizis et al.(2000)]{giz00}Gizis, J.\ E., Monet, D.\ G., Reid, I.\ N.,
Kirkpatrick, J.\ D., Liebert, J., \& Williams, R. 2000, \aj, 120, 1085

\bibitem[Goldman et al.(1999)]{gld99}Goldman, B., et al. 1999, \aap, 351,
L5

\bibitem[Goldreich \& Lynden-Bell(1969)]{gol69}Goldreich, P., \& Lynden-Bell, D. 1969, \apj, 156, 59

\bibitem[Golimowski et al.(2004)]{gol04} Golimowski, D.\ A., et al. 2004, \aj, 127, 3516


\bibitem[G\"{u}del(2002)]{gud02}G\"{u}del, M. 2002, \araa, 40, 217

\bibitem[G\"{u}del \& Benz(1993)]{gud93}G\"{u}del, M., \& Benz, A.\ O. 1993, \apj, 405, L63

\bibitem[Hall(2002a)]{hal02a}Hall, P.\ B. 2002a, \apj, 564, L89

\bibitem[Hall(2002b)]{hal02b}---. 2002b, \apj, 580, L77

\bibitem[Hambaryan et al.(2004)]{ham04}Hambaryan, V., Staude, A., Schwope, A.\ D.,
Scholz, R.-D., Kimeswenger, S., \& Neuh\"{a}user, R. 2004, \aap, 415, 265

\bibitem[Hambly et al.(2001)]{ham01}Hambly, N.\ C., et al. 2001, \mnras, 326, 1279

\bibitem[Hawley, Gizis, \& Reid(1996)]{haw96}Hawley, S.\ L., Gizis, J.\ E.,
\& Reid, I.\ N. 1996, \aj, 112, 2799

\bibitem[Ianna(1995)]{ian95}Ianna, P.\ A. 1995, in The Bottom of the Main Sequence
and Beyond, ed.\ C.\ G.\ Tinney (Heidelberg: Springer-Verlag), p.\ 138

\bibitem[Ip, Kopp, \& Hu(2004)]{ip04}Ip, W.-H., Kopp, A., \& Hu, J.-H. 2004, \apj, 602, L531

\bibitem[Jackson et al.(1987)]{jac87} Jackson, P.\ D., Kundu, M.\ R., \& White, S.\ M.
1987, \apj, 316, L85

\bibitem[Johns-Krull \& Valenti(1996)]{joh96} Johns-Krull, C.\ M., \& Valenti, J.\ A. 1996, \apj, 459, L95

\bibitem[Kastner et al.(1997)]{kas97}Kastner, J.\ H., Zuckerman, B., Weintraub, D.\ A.,
\& Forveille, T. 1997, Science, 277, 67

\bibitem[Katsova, Badalyan, \& Livshits(1987)]{kat87}Katsova, M.\ M., Badalyan, O.\ G.,
\& Livshits, M.\ A. 1987, Astron.\ Zh., 64, 1243

\bibitem[Kirkpatrick et al.(1999)]{kir99}Kirkpatrick, J.\ D., et al. 1999,
\apj, 519, 802

\bibitem[Krishnamurthi, Leto, \& Linksy(1999)]{kri99}Krishnamurthi, A., Leto, G.,
\& Linsky, J.\ L. 1999, \aj, 118, 1369

\bibitem[Kundu et al.(1987)]{kun87}Kundu, M.\ R., Jackson, P.\ D., White,
S.\ M., \& Melozzi, M. 1987, \apj, 312, 822

\bibitem[Lane, Boden, \& Kulkarni(2001)]{lan01}Lane, B.\ F., Boden, A.\ F.,
\& Kulkarni, S.\ R. 2001, \apj, 551, L81

\bibitem[Leggett et al.(2002)]{leg02} Leggett, S.\ K., Hauschildt, P.\ H.,
Allard, F., Geballe, T.\ R., \& Baron, E. 2002, \mnras, 332, 78

\bibitem[Leto et al.(2000)]{let00}Leto, G., Pagano, I., Linsky, J.\ L., Rodon{\'{o}}no, M.,
\& Umana, G. 2000, \aap, 359, 1035

\bibitem[Liebert et al.(2003)]{lie03}Liebert, J., Kirkpatrick, J.\ D., Cruz, K.\ L,
Reid, I.\ N., Burgasser, A.\ J., Tinney, C.\ G., \& Gizis, J.\ E.
2003, \aj, 125, 343

\bibitem[Liebert et al.(1999)]{lie99}Liebert, J., Kirkpatrick, J.\ D.,
Reid, I.\ N., \& Fisher,
M.\ D. 1999, \apj, 519, L345

\bibitem[Linsky \& Gary(1983)]{lin83}Linsky, J.\ L., \& Gary, D.\ E. 1983,
\apj, 274, 776

\bibitem[Lodieu, Scholz, \& McCaughrean(2002)]{lod02} Lodieu, N., Scholz, R.-D., \&
McCaughrean, M.\ J. 2002, \aap, 389, L20

\bibitem[Luyten(1979a)]{luy79a}Luyten, W.\ J. 1979a, LHS Catalogue: A Catalogue of Stars with Proper
Motions Exceeding 0$\farcs$5 Annually (Minneapolis: Univ.\ Minn.\ Press)

\bibitem[Mart{\'{\i}}n \& Ardila(2001)]{mrt01}Mart{\'{\i}}n, E.\ L.,
\& Ardila, D.\ R. 2001, \aj, 121, 2758


\bibitem[Meyer \& Meyer-Hofmeister(1999)]{mey99}Meyer, F., \& Meyer-Hofmeister, E. 1999, \aap, 341, L23

\bibitem[Mohanty \& Basri(2003)]{moh03}Mohanty, S., \& Basri, G. 2003, \apj, 583, 451

\bibitem[Mohanty et al.(2002)]{moh02}Mohanty, S., Basri, G., Shu, F., Allard, F., \& Chabrier, G. 2002,
\apj, 572, 469

\bibitem[Neuh\"{a}user et al.(1999)]{neu99}Neuh\"{a}user, R., et al. 1999, \aap, 343, 883

\bibitem[Neuh\"{a}user et al.(2002)]{neu02}Neuh\"{a}user, R., et al. 2002, AN, 323, 447

\bibitem[Noyes et al.(1984)]{noy84}Noyes, R.\ W., Hartmann, L.\ W., Baliunas, S.\ L., Duncan, D.\ K.,
\& Vaughan, A.\ H. 1984, \apj, 279, 763

\bibitem[Pallavicini et al.(1981)]{pal81}Pallavicini, R., Golub, L., Rosner, R.,
Vaiana, G.\ S., Ayres, T., \& Linsky, J.\ L. 1981, \apj, 248, 279

\bibitem[Paresce(1984)]{par84}Paresce, F. 1984, \aj, 89, 1022

\bibitem[R{\"{a}}dler et al.(1990)]{rad90} R{\"{a}}dler, K.\ -H., Wiedemann, E., Bradenberg, A.,
Meinel, R., \& Tuominen, I. 1990, \aap, 239, 413

\bibitem[Razin(1960)]{raz60} Razin, V.\ A. 1960, Radiofiz., 3, 584

\bibitem[Rebolo, Mart{\'{\i}}n, \& Magazzu(1992)]{reb92}Rebolo, R.,
Mart{\'{\i}}n, E.\ L., \& Magazzu, A. 1992, \apj, 389, L83

\bibitem[Reid \& Hawley(2000)]{rei00a}Reid, I.\ N., \& Hawley, S.\ L.
2000, New Light on Dark Stars (Chichester: Praxis)

\bibitem[Reid, Hawley, \& Mateo(1995)]{rei95}Reid, I.\ N., Hawley, S.\ L.,
\& Mateo, M. 1995, \mnras, 272, 828

\bibitem[Reid et al.(1999)]{rei99}Reid, I.\ N., Kirkpatrick,
J.\ D., Gizis, J.\ E., \& Liebert, J. 1999, \apj, 527, L105

\bibitem[Richards et al.(1998)]{ric98}Richards, E.\ A., Kellermann, K.\ I., Fomalont, E.\ B.,
Windhorst, R.\ A., \& Partridge, R.\ B. 1998, \aj, 116, 1039

\bibitem[Rutledge et al.(2000)]{rut00}Rutledge, R.\ E., Basri, G.,
Mart{\'{\i}}n, E.\ L., \& Bildsten, L. 2000, \apj, 538, L141

\bibitem[Ruiz et al.(1990)]{rui90}Ruiz, M.\ T., Anguita, C., Maza, J., \& Roth, M.
1990, \aj, 100, 1270

\bibitem[Saar \& Linsky(1985)]{saa85}Saar, S.\ H., \& Linsky, J.\ L. 1985, \apj, 299, L47

\bibitem[Schmitt, Fleming, \& Giampapa(1995)]{sch95}Schmitt, J.\ H.\ M.\ M.,
Fleming, T.\ A., \& Giampapa, M.\ S. 1995, \apj, 450, 392

\bibitem[Schmitt \& Liefke(2002)]{scm02} Scmitt, J.\ H.\ M.\ M., \& Liefke, C. 2002, \aap, 382, L9

\bibitem[Schmitt \& Liefke(2004)]{scm04} ---. 2004, \aap, 417, 651

\bibitem[Scholz \& Meusinger(2002)]{sch02} Scholz, R.-D., \& Meusinger, H. 2002, \mnras, 336, L49

\bibitem[Segransan et al.(2003)]{seg03} S\'{e}gransan, D., Kervella, P., Forveille, T.,
\& Queloz, D. 2003, \aap, 397, L5

\bibitem[Stepanov et al.(2001)]{ste01} Stepanov, A.\ V., Kliem, B., Zaitsev, V.\ V., F\"{u}rst, E.,
Jessner, A., Kr\"{u}ger, A., Hildebrandt, J., \& Schmitt, J.\ H.\ M.\ M. 2001, \aap, 374, 1072

\bibitem[Stevenson(2003)]{ste03} Stevenson, D.\ J. 2003, Earth Planet.\ Sci.\ Lett., 208, 1

\bibitem[Tinney(1996)]{tin96} Tinney, C.\ G. 1996, \mnras, 281, 644

\bibitem[Tinney \& Reid(1998)]{tin98}Tinney, C.\ G., \& Reid, I.\ N. 1998, \mnras, 301, 1031

\bibitem[Tsytovich(1951)]{tsy51} Tsytovich, V.\ N. 1951, Vestn.\ Mosk.\ Univ.\ Phys., 11, 27

\bibitem[van Altena, Lee, \& Hoffleit(1995)]{van95}van Altena, W.\ F., Lee,
J.\ T., \& Hoffleit, E.\ D. 1995, The General Catalog of Trignometric
Stellar Parallaxes, 4$^{th}$ Edition (New Haven: Yale Univ.\ Obs.)

\bibitem[van den Besselaar et al.(2003)]{van03}van den Besselaar, E.\ J.\ M., Raassen, A.\ J.\ J., Mewe, R.,
van der Meer, R.\ L.\ J., G\"{u}del, M., \& Audard, M. 2003, \aap, 411, 587

\bibitem[West et al.(2004)]{wes04}West, A.\ A., et al. 2004, \aj, 128, 426

\bibitem[White, Jackson, \& Kundu(1989)]{whi89}White, S.\ M., Jackson, P.\ D., \&
Kundu, M.\ R. 1989, \apjs, 71, 895

\bibitem[Winglee, Dulk, \& Bastian(1986)]{win86}Winglee, R.\ M., Dulk, G.\ A.,
\& Bastian, T.\ S. 1986, \apj, 309, L59



\end{thebibliography}
\end{document}